\documentclass[journal]{IEEEtran}[12pt]
\ifCLASSINFOpdf
\usepackage[pdftex]{graphicx}
\graphicspath{{../pdf/}{../jpeg/}}
\DeclareGraphicsExtensions{.pdf,.jpeg,.png}
\else
\usepackage[dvips]{graphicx}
\graphicspath{{../eps/}}
\DeclareGraphicsExtensions{.eps}
\fi\usepackage{graphicx}

\usepackage{graphics}
\usepackage{epsfig}
\usepackage{epstopdf}
\usepackage{stfloats}
\usepackage[cmex10]{amsmath}
\usepackage{algorithmic}
\usepackage[ruled]{algorithm2e}
\usepackage{indentfirst}
\usepackage{array}
\usepackage{mdwmath}
\usepackage{mdwtab}
\usepackage{graphicx}
\usepackage{subfigure}
\usepackage{color}
\usepackage{amsfonts,amssymb}
\usepackage{multirow}
\usepackage{diagbox}
\usepackage{caption}
\usepackage{verbatim}
\usepackage{float}
\usepackage{graphicx}
\usepackage{bm}
\usepackage{bbding}
\usepackage{tabu} 
\usepackage{multirow} 
\usepackage{multicol} 
\usepackage{multirow} 
\usepackage{float} 
\usepackage{makecell}
\usepackage{booktabs} 
\usepackage [mathscr] {euscript}
\usepackage{threeparttable}
\usepackage{hyperref} %

\usepackage{lineno} 

\usepackage{amsfonts,amsthm,array} 

\usepackage{amsfonts,amsthm,array} 
\makeatletter
\renewcommand{\maketag@@@}[1]{\hbox{\m@th\normalsize\normalfont#1}}
\makeatother

\begin{document}
	

\title{Trade-off for Secure UAV-ISCC Systems \thanks{Manuscript received.}}

\author{Hongjiang~Lei, 
	Jun~He,
	Congke~Jiang,
	Ki-Hong~Park,\\  
	Wenqian~Shen, 
	Liang~Yang, 
	and 
	Gaofeng~Pan 
\thanks{Hongjiang~Lei, Jun~He, and Congke~Jiang are with School of Communications and Information Engineering, Chongqing University of Posts and Telecommunications, Chongqing 400065, China  (e-mail: leihj@cqupt.edu.cn, cqupthj@163.com, cquptjck@163.com).}
\thanks{Ki-Hong~Park is with CEMSE Division, King Abdullah University of Science and Technology (KAUST), Thuwal 23955-6900, Saudi Arabia (e-mail: kihong.park@kaust.edu.sa).}
\thanks{Wenqian~Shen is with School of Information and Electronics, Beijing Institute of Technology, Beijing 100081, China (e-mail: shenwq@bit.edu.cn).}
\thanks{Liang~Yang is with College of Computer Science and Electronic Engineering, Hunan University, Changsha 410082, China	(e-mail: liangy@hnu.edu.cn).}	
\thanks{Gaofeng~Pan is with School of Cyberspace Science and Technology, Beijing Institute of Technology, Beijing 100081, China (e-mail: gfpan@bit.edu.cn).}	
}

\maketitle
\begin{abstract}
	
The integrated sensing, communication, and computing (ISCC) system overcomes the limitations of conventional standalone architectures. Through resource sharing and collaborative design, it dynamically optimizes and jointly enhances communication, sensing, and computing performance, thereby significantly improving overall system efficiency. This work investigates the performance trade-off among secure communication rate, radar estimation rate, and computational energy efficiency in an uncrewed aerial vehicle (UAV)-assisted ISCC system. By jointly optimizing the UAV’s three-dimensional (3D) trajectory, beamforming, user scheduling, and computational frequency, three optimization problems are formulated to maximize the average secrecy rate, sensing rate, and computational energy efficiency, respectively, thus establishing the system’s performance boundaries under diverse scenarios. On this basis, the trade-off among security, sensing, and computation is further explored with the goal of maximizing the normalized weighted sum of the three performance metrics, which provides a theoretical basis for the performance-coordinated design of aerial ISCC systems.

\end{abstract}


\section{Introduction}
\label{sec:Introduction}
\subsection{Background and Related Works}

Integrated sensing and communication (ISAC) leverages the similarities between communication systems and sensing systems in terms of channel characteristics, hardware equipment, and signal processing workflows. Through unified waveform design and the sharing of software and hardware resources, it enables the collaborative operation of communication and sensing functions, thereby improving spectral efficiency while reducing hardware costs and power consumption \cite{ZhangD2026Surveys}. Building upon ISAC, the integrated sensing, communication, and computing (ISCC) system further incorporates computing capabilities, such as mobile edge computing (MEC), into a unified framework, establishing a new paradigm  \cite{WangX20024ACS}. Specifically, on the foundation of resource sharing between communication and sensing achieved by ISAC, the ISCC system introduces MEC, which pushes computing power to the network edge. This allows sensing data to be processed locally, avoiding the high latency and additional energy consumption associated with uploading data to the cloud \cite{LiC2026Surveys, WenD2025CST}.

The multifunctional ISCC framework presented a rich landscape for research, amenable to exploration across several distinct optimization objectives, which were broadly categorized into three types: single-performance maximization, dual-performance optimization, and multi-performance trade-off. Among the single-performance maximization works, Ref. \cite{ZhaoY2025TCOM} considered a terrestrial ISCC system and maximized the Cramér–Rao bound (CRB) through the joint optimization of beamforming design and computational frequency, while Ref. \cite{WangZ2023JSAC} investigated a multi-functional base station and maximized the computation rate by optimizing computational resources and beamforming strategies. For dual-performance optimization, Ref. \cite{CangY2025JSAC} studied a multi-static ISCC system and formulated the energy consumption minimization problem via beamforming design. Ref. \cite{DingY2023JSTSP} maximized secure computation efficiency by jointly optimizing offloading decisions, time allocation, local computing bits, and transmit power. And Ref. \cite{DingC2022JSAC} maximized the trade-off between radar sensing and data offloading through the joint design of beamforming and computing resources. 
For multi-performance trade-off, Ref. \cite{SunG2024TVT} maximized the weighted sum of communication rate, sensing rate, and computing rate by jointly optimizing beamforming, resource allocation, and offloading strategies.

Uncrewed aerial vehicles (UAVs) leverage their maneuverability and rapid deployment to provide line-of-sight (LoS) links and multi-angle sensing from an aerial perspective, effectively compensating for ground sensor limitations while enabling local processing and avoiding high latency and energy consumption, thereby meeting the stringent requirements of autonomous driving and emergency communications \cite{WangY2025TCCN}. Several recent studies have explored UAV-ISCC systems from various perspectives. Refs. \cite{ChenJ2024WCL} and \cite{ZhouY2024IoT} proposed single-UAV-enabled frameworks, where a multi-functional UAV performed sensing, edge computing, and communication tasks; the former maximized computing efficiency by jointly optimizing two-dimensional (2D) trajectory, beamforming, and offloading strategy under a sensing threshold, while the latter minimized weighted total energy consumption via joint optimization of UAV central processing unit (CPU) frequency, radar sensing power, user transmit power, and 2D trajectory. Ref. \cite{XuY2023WCL} considered a UAV-ISCC framework, where a multi-functional UAV performed local computation and offloaded its computational tasks to ground access points while simultaneously sensing a target using unified beam waves; the Pareto boundary between computation capacity and sensing beampattern gain was characterized to unveil the fundamental trade-off between the two functions. Ref. \cite{VanChienT2024CL} investigated joint computation offloading and target tracking in an ISAC-enabled UAV network, where the UAV partially offloaded computing tasks to a ground user while simultaneously using the offloading bit sequence to estimate target velocity via an autocorrelation function, minimizing computation latency and the CRB for velocity estimation under budget constraints. Refs. \cite{XuS2025TMC} and \cite{ZhouY2026TGCN} both considered multi-UAV or relay-assisted configurations with MEC; the former studied joint trajectory and beamforming design for UAV-relayed ISAC systems in clutter environments, adopting a relay-based ISAC-then-offload frame structure to maximize throughput while ensuring sensing accuracy. The latter proposed a multi-UAV ISCC system where multiple UAVs offloaded sensing data to a high-altitude platform via MEC, formulating a trade-off between sensing data acquisition and total energy consumption. 
Ref. \cite{HuangN2023IoT} aimed to minimize a system-wide cost that accounted for both the UAV's energy consumption and the data collection time, subject to constraints on model training error and sensing performance, and formulated a joint optimization problem involving sensing scheduling, the number of time slots, power allocation, and UAV trajectory. 
Ref. \cite{PengS2025TCOM} further extended this paradigm by proposing a multi-UAV-assisted ISCC network, in which multi-functional UAVs provided communication and edge computing services to mobile users while performing target sensing using multiple-input multiple-output (MIMO) arrays, with joint optimization of transmit beamforming, UAV trajectories, compression ratios, offloading partitions, and computation resource allocation under sensing quality constraints. Collectively, these works highlight the growing research interest in UAV-ISCC systems, with a shared emphasis on energy efficiency, computation latency, and sensing accuracy through joint optimization of trajectory, resource allocation, and offloading strategies.

The integration of UAVs with MEC enhances system flexibility and efficiency but introduces security risks such as data eavesdropping due to the broadcast nature of wireless channels. Physical layer security and covert communication utilize beamforming and artificial noise techniques to provide a solid foundation for secure UAV-ISCC systems. Ref. \cite{LeiH2025IoTCongke} considered a secure UAV-ISCC system, where the UAV transmitted radar signals to locate and suppress an eavesdropper while providing offloading services to ground users. Under constraints on UAV speed, power, propulsion energy, data transmission, and computation time, the authors minimized ground user energy consumption by jointly optimizing offloading ratio, scheduling, beamforming, and UAV trajectory. Ref. \cite{ZhangQ2026JSAC} proposed a novel covert transmission scheme based on an active reconfigurable intelligent surface (RIS)-enabled UAV-ISCC framework, where the multi-functional UAV realized simultaneous target sensing and uplink covert communication while performing edge computing for users. To maximize the minimum covert transmission rate among all uplink users, the authors jointly designed UAV transmit beamforming and trajectory, RIS weights, power allocation, and signal processing in the full-duplex uplink transmission system. Collectively, these works highlight growing interest in securing ISCC systems, with a shared focus on energy efficiency, latency, and sensing accuracy through joint resource optimization. 

\subsection{Motivation and Contributions}

The existing research on UAV-ISCC has expanded from single-performance optimization to dual-performance optimization and multi-performance trade-offs. However, most studies have focused on traditional performance such as communication rate, sensing accuracy, or energy consumption, while research on the security performance of UAV-ISCC remains scarce and is mostly limited to single objectives. To address these gaps, this paper focuses on UAV-ISCC systems by jointly optimizing the UAV's three-dimensional (3D) trajectory, beamforming, resource scheduling, and computational frequency to maximize the secure communication rate, sensing rate, and computational energy efficiency, respectively. A normalized weighted sum method is then employed to achieve a trade-off among the three metrics, providing theoretical support for the coordinated design of security, sensing, and computation in UAV-ISCC systems.
The main contributions of this paper are summarized as follows.
\begin{enumerate}
	\item This work investigates a UAV-ISCC system, where a rotorcraft UAV provides secure communication and sensing services to ground users and targets, while performing local computation or offloading tasks to a ground base station, with an eavesdropper attempting to intercept the transmissions. To comprehensively evaluate system performance, three optimization problems are first formulated to maximize the secure communication rate, sensing rate, and computational energy efficiency, respectively, through the joint optimization of the UAV’s 3D trajectory, beamforming, resource scheduling, and computational frequency. These formulations establish the performance boundaries of the system under different application scenarios. To address the inherent inconsistency in the dimensions of the three performance metrics, a multi-performance trade-off problem is further developed to maximize their normalized weighted sum, thereby achieving coordinated system design of security, sensing, and computation. The resulting non-convex optimization problems are effectively solved by employing block coordinate descent and successive convex approximation (SCA) techniques, leading to the design of an efficient iterative algorithm.

	\item Although several outstanding works, such as Refs. \cite{DingC2022JSAC} and \cite{SunG2024TVT}, have investigated the trade-off between sensing and data offloading or maximized the weighted sum of communication rate, sensing rate, and computing rate, they are primarily limited to terrestrial scenarios without considering security threats. In contrast, this work extends the ISCC system to an aerial scenario by introducing a UAV and further incorporates physical layer security into the framework. The secure communication rate is adopted as a core performance metric, optimized jointly with the sensing rate and computational energy efficiency. To address the inconsistency in dimensions among the secure rate, sensing mutual information, and computational energy efficiency, a normalized weighted sum method is proposed to achieve a balanced trade-off among the three metrics. Moreover, the UAV's 3D trajectory is introduced as a new optimization variable, jointly optimized with beamforming, resource scheduling, and computational frequency.

\end{enumerate}

\subsection{Organization}
The remainder of this paper is organized as follows. Section \ref{sec:SystemModel} presents the system model. Sections \ref{sec:CommunicationCen}, \ref{sec:SensingCen}, and \ref{sec:ComputingCen} formulate the optimization problems and present their solutions for communication-centric, sensing-centric, and computing-centric energy efficiency scenarios, respectively. Section \ref{sec:WeightedOptimization} introduces the normalized weighted optimization problem and provides its solution. Section \ref{sec:Simulation} presents the simulation results, and Section \ref{sec:Conclusion} concludes the paper.

\section{System Model and Problem Formulation}
\label{sec:SystemModel}

\begin{figure}
	\centering
	\includegraphics[width = 0.4 \textwidth]{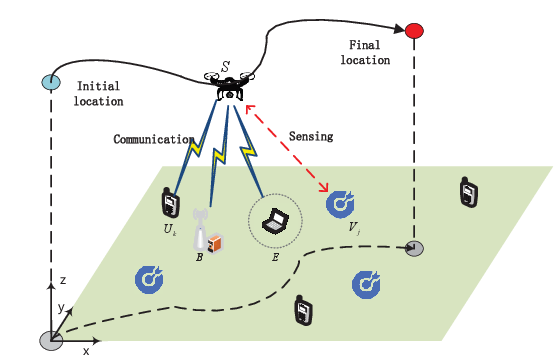}
	\caption{System model of a secure UAV-aided ISCC system.}
	\label{figmodel}
\end{figure}

As illustrated in Fig.~\ref{figmodel}, the considered system consists of a rotorcraft UAV ($S$), multiple ground communication users (denoted as $U_k$, $k = \{1, \ldots, K\}$), a base station $B$ equipped with an edge server, and multiple sensing targets (denoted as $V_j$ with $j \in \mathcal{J} = \{1, \ldots, J\}$). Additionally, there exists an unknown and uncertain eavesdropper ($E$) that attempts to intercept the communication signals of the ground users. The total flight time of the UAV is denoted by $T$, which is equally divided into $N$ time slots, each of duration $\delta_t = T/N$. In each time slot, $S$ first performs sensing toward the targets and then conducts downlink communication with the ground users. During the communication phase, part or all of the received sensing data is transmitted to the base station $B$ for further processing and analysis.

By establishing a 3D Cartesian coordinate system, the horizontal coordinates of $U_k$, $V_j$, $B$, $E$, and $S$ are denoted as $\mathbf{q}_{U_k} = [x_{U_k}, y_{U_k}]^T$, $\mathbf{q}_{V_j} = [x_{V_j}, y_{V_j}]^T$, $\mathbf{q}_B = [x_B, y_B]^T$, $\mathbf{q}_E = [x_E, y_E]^T$, and $\mathbf{q}_s\left[ n \right] = [x_s\left[ n \right], y_s\left[ n \right]]^T$, respectively, with $z_s\left[ n \right]$ representing the altitude of $S$ in the $n$-th time slot. 
All ground users are equipped with a single antenna, while $S$ is equipped with two uniform planar arrays (UPAs) for signal transmission and reception, each comprising $M = M_x \times M_y$ antenna elements, where $M_x$ and $M_y$ denote the numbers of elements along the $x$- and $y$-axes, respectively, with adjacent elements spaced by half a wavelength. 

The air-to-ground channels are characterized by a probabilistic LoS (PLoS) model. In the $n$-th time slot, the LoS probability between $S$ and a ground node $l \in \mathcal{L} = \{U_1, \ldots, U_K, B, E, V_1, \ldots, V_J\}$ is given by  
$P_l^{\mathrm{L}}\left[ n \right] = \left( 1 + C \exp\left( -D \left( \theta_l\left[ n \right] - C \right) \right) \right)^{-1}$, where $C > 0$ and $D > 0$ are environment-dependent constants \cite{AlHouraniA20217WCL}. The elevation angle $\theta_l\left[ n \right]$ is computed as  
$\theta_l\left[ n \right] = \frac{180}{\pi} \arctan\left( \frac{z_s\left[ n \right]}{d_{h,l}\left[ n \right]} \right)$, 
with $d_{h,l}\left[ n \right] = \| \mathbf{q}_s\left[ n \right] - \mathbf{q}_l \|$ denoting the horizontal distance. Accordingly, the non-LoS (NLoS) probability is obtained as $P_l^{\mathrm{N}}\left[ n \right] = 1 - P_l^{\mathrm{L}}\left[ n \right]$ \cite{LeiH2025TCCN}.
In the same time slot, the downlink communication channel from $S$ to a ground node $c \in \mathcal{C} = \{U_1, \ldots, U_K, B, E\}$ is modeled as  
\begin{align}
	\mathbf{h}_c\left[ n \right] = 
	\left\{ {\begin{array}{*{20}{c}}
			{\sqrt {\frac{{{\beta _0}}}{{d_c^2\left[ n \right]}}} {\bf{a}}_c^H\left[ n \right],}&{{\rm{LoS}},}\\
			{\sqrt {\frac{{\alpha {\beta _0}}}{{d_c^2\left[ n \right]}}} {\bf{a}}_{{\rm{NLoS}}}^H\left[ n \right],}&{{\rm{NLoS}},}
	\end{array}} \right.
	\label{hc}
\end{align}
where $\beta_0$ denotes the reference channel gain at unit distance, and $d_c\left[ n \right] = \sqrt{z_s^2\left[ n \right] + d_{h,c}^2\left[ n \right]}$ is the Euclidean distance between $S$ and node $c$, $\alpha < 1$ accounts for the additional attenuation under NLoS conditions. 
The NLoS component $\mathbf{a}_{\text{NLoS}}^H\left[ n \right]$ is modeled as a complex Gaussian random vector with zero mean and unit covariance matrix \cite{DengC2023TWC}.
The steering vector $\mathbf{a}_l^H\left[ n \right]$ for a ground node $l$ is defined as  
$\mathbf{a}_l^H\left[ n \right] = \left[ 1, e^{j\pi \Theta_l^x\left[ n \right]}, \ldots, e^{j\pi (M_x-1) \Theta_l^x\left[ n \right]} \right] \otimes \left[ 1, e^{j\pi \Theta_l^y\left[ n \right]}, \ldots, e^{j\pi (M_y-1) \Theta_l^y\left[ n \right]} \right]$, 
with directional cosines $\Theta_l^x\left[ n \right] = \frac{x_s\left[ n \right] - x_l}{d_l\left[ n \right]}$ and $\Theta_l^y\left[ n \right] = \frac{y_s\left[ n \right] - y_l}{d_l\left[ n \right]}$ \cite{MengK2023TWC}, where $\otimes$ represents the Kronecker product.
For sensing purposes, the signals transmitted by $S$ are reflected by the targets and then received back by $S$. It is assumed that the Doppler shifts caused by the motion of both UAVs and the targets remain constant within a single time slot and is effectively compensated \cite{LiuF2018TSP}. The sensing channel associated with target $V_j$ is expressed as \cite{JingX2024TWC}  
\begin{align}
	\mathbf{h}_{V_j}\left[ n \right] = 
	\left\{ {\begin{array}{*{20}{c}}
			{\sqrt {\frac{{{\beta _0}}}{{d_{{V_j}}^{ 4}\left[ n \right]}}} {\bf{a}}_{{V_j}}\left[ n \right]{\bf{a}}_{{V_j}}^H\left[ n \right],}&{{\rm{LoS}},}\\
			{\sqrt {\frac{{\alpha {\beta _0}}}{{d_{{V_j}}^{4}\left[ n \right]}}} {\bf{a}}_{{\rm{NLoS}}}\left[ n \right]{\bf{a}}_{{\rm{NLoS}}}^H\left[ n \right],}&{{\rm{NLoS}}.}
	\end{array}} \right.
	\label{hvj}
\end{align}

Binary scheduling variables $\theta_{U_k}\left[ n \right]$ and $\theta_{V_j}\left[ n \right]$ are introduced to indicate whether $U_k$ is scheduled for communication with $S$ and whether $V_j$ is sensed by $S$ in the $n$-th time slot, respectively. Specifically, $\theta_{U_k}\left[ n \right] = 1$ if $U_k$ is selected for communication, and $\theta_{V_j}\left[ n \right] = 1$ if $V_j$ is selected for sensing \cite{ZhangJ2024TWC}. It is assumed that one user and one target is served at most once during the flight period, the following constraints are imposed:
\begin{subequations}
\begin{align}
	&\sum_{n=1}^{N} \theta_{U_k}\left[ n \right] \leq 1, \forall n, \label{scheduling1a}\\
	&\theta_{U_k}\left[ n \right] \in \{0, 1\}, \forall n, k,  \label{scheduling1b}\\
	&\sum_{n=1}^{N} \theta_{V_j}\left[ n \right] \leq 1, \forall n, \label{scheduling2a} \\	
	&\theta_{V_j}\left[ n \right] \in \{0, 1\}, \forall n, j. \label{scheduling2b}
\end{align}
\end{subequations}

\subsection{Communication and Sensing Model}
In the $n$-th time slot, the signal transmitted by $S$ is given by
\begin{align}
	\mathbf{x}\left[ n \right] &= \mathbf{w}_K\left[ n \right] \sum_{k=1}^{K} \theta_{U_k}\left[ n \right] s_{U_k}\left[ n \right] \notag \\
	& + \mathbf{w}_B\left[ n \right] s_B\left[ n \right] + \mathbf{w}_J\left[ n \right] \sum_{j=1}^{J} \theta_{V_j}\left[ n \right] s_{V_j}\left[ n \right],
	\label{H232}
\end{align}
where 
$\mathbf{w}_K\left[ n \right] \in \mathbb{C}^{M \times 1}$, $\mathbf{w}_B\left[ n \right] \in \mathbb{C}^{M \times 1}$, and $\mathbf{w}_J\left[ n \right] \in \mathbb{C}^{M \times 1}$ denote the transmit beamforming vectors for communication, offloading transmission, and sensing, respectively. 
The signal received by ground node $c$ is expressed as $y_c \left[ n \right] = {{\bf{h}}_c}\left[ n \right]{\bf x}\left[ n \right] + n_c$, where ${n_c} \sim CN\left( {0,\sigma _c^2} \right)$ represents the additive white Gaussian noise (AWGN) at ground node $c$.

The signal-to-interference-plus-noise ratios (SINRs) for the legitimate user $U_k$, the eavesdropper $E$, and the base station $B$ are then given by
\begin{align}
	\gamma_{U_k}^{\mathrm{L}}\left[ n \right] &= \frac{ \theta_{U_k}\left[ n \right] \left| \mathbf{a}_{U_k}^H\left[ n \right] \mathbf{w}_K\left[ n \right] \right|^2 }{ \left| \mathbf{a}_{U_k}^H\left[ n \right] \mathbf{w}_J\left[ n \right] \right|^2 + \left| \mathbf{a}_{U_k}^H\left[ n \right] \mathbf{w}_B\left[ n \right] \right|^2 + \mu_{U_k}\left[ n \right] }, \\
	\gamma_{E_k}^{\mathrm{L}}\left[ n \right] &= \frac{ \theta_{U_k}\left[ n \right] \left| \mathbf{a}_E^H\left[ n \right] \mathbf{w}_K\left[ n \right] \right|^2 }{ \left| \mathbf{a}_E^H\left[ n \right] \mathbf{w}_J\left[ n \right] \right|^2 + \left| \mathbf{a}_E^H\left[ n \right] \mathbf{w}_B\left[ n \right] \right|^2 + \mu_E\left[ n \right] }, \\
	\gamma_B^{\mathrm{L}}\left[ n \right] &= \frac{ \left| \mathbf{a}_B^H\left[ n \right] \mathbf{w}_B\left[ n \right] \right|^2 }{ \left| \mathbf{a}_B^H\left[ n \right] \mathbf{w}_{J}\left[ n \right] \right|^2 + \left| \mathbf{a}_B^H\left[ n \right] \mathbf{w}_{K}\left[ n \right] \right|^2 + \mu_B\left[ n \right] },
\end{align}
where 
$\mu_c\left[ n \right] = \beta_0^{-1} \sigma_c^2 d_c^2\left[ n \right]$.
The achievable rate at node $c \in \{U_k, E, B\}$ in the $n$-th time slot is expressed as
\begin{align}
	R_c\left[ n \right] &= P_c^{\mathrm{L}}\left[ n \right] \log_2 \left( 1 + \gamma_c^{\mathrm{L}}\left[ n \right] \right) \nonumber \\
	&+ P_c^{\mathrm{N}}\left[ n \right] \log_2 \left( 1 + \gamma_c^{\mathrm{N}}\left[ n \right] \right),
	\label{ratec}
\end{align}
where $\gamma_c^{\mathrm{N}}\left[ n \right]$ denotes the SINRs with NLoS conditions. Since the achievable rate under NLoS propagation is significantly lower than that under LoS propagation, it is approximated by its LoS component \cite{LeiH2024TCCN3D}\footnote{
	A simple approach is to consider the worst-case scenario, i.e.,  $R_{E_k}^{\mathrm{ub}}\left[ n \right] = \log_2 \left( 1 + \gamma_{E_k}^{\mathrm{L}}\left[ n \right] \right)$. It should be noted that the method utilized in this work is more accurate and more challenging since $P_c^{\mathrm{L}}\left[ n \right]$ depends on the trajectory of UAVs. 	
}:
\begin{align}
	R_c^{\mathrm{app}}\left[ n \right] = P_c^{\mathrm{L}}\left[ n \right] \log_2 \left( 1 + \gamma_c^{\mathrm{L}}\left[ n \right] \right). \label{rate_lb}
\end{align}

Finally, the achievable secrecy rate for $U_k$ is given by
\begin{align}
	R_{U_k}^{\mathrm{sec}}\left[ n \right] = \left[ R_{U_k}^{\mathrm{app}}\left[ n \right] - R_{E_k}^{\mathrm{app}}\left[ n \right] \right]^+,
	\label{H232}
\end{align}
where $[x]^+ \triangleq \max(x, 0)$.

Similarly, considering only the reflected signals received via the LoS channel, the sensing SINR for $V_j$ at $S$ is expressed as
\begin{align}
	\gamma_{V_j}\left[ n \right] = \frac{ \theta_{V_j}\left[ n \right] \left\| \mathbf{h}_{V_j}\left[ n \right] \mathbf{w}_J\left[ n \right] \right\|^2 }{ \sigma_s^2 },
	\label{gamma_vj}
\end{align}
where $\sigma_s^2$ denotes the noise power at the UAV receiver.

Following \cite{HuangN2023IoT, ChiriyathARTSP2016}, the radar estimation rate is utilized as the performance metric to quantify the amount of target-related information that the ISAC device can extract from the received echo signal. In the considered system, the radar estimation rate for $V_j$ is given by
\begin{align}
	R_{V_j}^{\mathrm{rad}}\left[ n \right] = \frac{\delta}{2\mu} P_{V_j}^{\mathrm{L}}\left[ n \right] \log_2 \left( 1 + 2 \sigma_{\mathrm{pre}}^2 \hat{\gamma}^2 B_w^3 \mu \, \gamma_{V_j}\left[ n \right] \right),
	\label{Rrad}
\end{align}
where $\delta$ is the radar duty factor, $\mu$ denotes the radar pulse duration, $\hat{\gamma}$ is a constant determined by the shape of the radar waveform, $B_w$ represents the bandwidth, and $\sigma_{\mathrm{pre}}^2$ is the variance of the predicted radar return.

\subsection{Computing Model}

The onboard ISAC device $S$ performs radar sensing and the received echo signal is converted into a set of data bits. The amount of sensing data (in bits) generated for $V_j$ in one time slot is denoted by $D_J$, which is modeled as a constant determined by the radar sensing configuration \cite{DingC2022JSAC}. Due to limited onboard resources, $S$ may offload a portion of the sensed data to the edge server located at $B$ for processing. The total amount of data that $S$ can process in the $n$-th time slot is given by
\begin{align}
	D_s\left[ n \right] = \delta_tR_B^{\mathrm{app}}\left[ n \right] B_w + \frac{f_s\left[ n \right]}{F_s},
	\label{Dsn}
\end{align}
where $F_s$ (cycles/bit) denotes the required CPU cycles per bit, and $f_s\left[ n \right]$ (cycles/s) represents the local computing capacity allocated at $S$ during the $n$-th time slot. The first term in (\ref{Dsn}) corresponds to the amount of data offloaded to $B$ for processing, while the second term corresponds to the amount of data processed locally. It is assumed that $B$ has sufficient energy and computational resources to process the received data; therefore, the energy consumption and processing delay at $B$ are not considered in this work.

The total energy consumption of $S$ is expressed as
\begin{align}
	E^{\mathrm{UAV}} = \sum_{n=1}^{N} \left( E^{\mathrm{cop}}\left[ n \right] + E^{\mathrm{tra}}\left[ n \right] + E^{\mathrm{fly}}\left[ n \right]  \right),
	\label{E_UAV}
\end{align}
where 
$E^{\mathrm{cop}}\left[ n \right] = \delta_t \kappa f_s^3\left[ n \right]$
denotes the computing energy consumption at $S$ \cite{LuW2022TCOM}, 
$E^{\mathrm{tra}}\left[ n \right] = \delta_t P^{\mathrm{tra}}\left[ n \right]$
denotes the transmission-related energy consumption, 
and 
$E^{\mathrm{fly}}\left[ n \right] = {\delta _t}\left\{ {{P^{{\rm{hor}}}}\left[ n \right] + {P^{{\rm{ver}}}}\left[ n \right]} \right\}$ 
denotes the propulsion energy consumption of $S$ in the horizontal and vertical directions. 
$P^{\mathrm{tra}}\left[ n \right] $ is expressed as
\begin{align}
	P^{\mathrm{tra}}\left[ n \right] &= {\rm{tr}}\left( {{{ {{{\bf{W}}_K}\left[ n \right]} }}} \right) + {\rm{tr}}\left( {{{ {{{\bf{W}}_J}\left[ n \right]} }}} \right) + {\rm{tr}}\left( {{{ {{{\bf{W}}_B}\left[ n \right]} }}} \right),
	\label{H232}
\end{align}
where $\mathbf{W}_i\left[ n \right] = \mathbf{w}_i\left[ n \right] \mathbf{w}_i^H\left[ n \right]$ for $i \in \{K, J, B\}$. 
The propulsion power is modelled as \cite{ZengY2019TWC, LeiH2025TAES}
\begin{subequations}
\begin{align}
	P^{\mathrm{hor}}\left[ n \right] &= P_{\mathrm{i}} \left( \sqrt{1 + \frac{V_{\mathrm{h}}^4\left[ n \right]}{4v_0^4}} - \frac{V_{\mathrm{h}}^2\left[ n \right]}{2v_0^2} \right)^{\frac{1}{2}} + \frac{1}{2} d_0 \rho s A V_{\mathrm{h}}^3\left[ n \right] \nonumber \\
	&\quad + P_{\mathrm{b}} \left( 1 + \frac{3 V_{\mathrm{h}}^2\left[ n \right]}{U_{\mathrm{tip}}^2} \right), \label{P_hor} \\
	P^{\mathrm{ver}}\left[ n \right] &= W V_z\left[ n \right], \quad \forall V_{z}\left[ n \right]>0,
\end{align}
\end{subequations}
where 
$V_{\mathrm{h}}\left[ n \right] = \frac{\| \mathbf{q}_s\left[ n + 1 \right] - \mathbf{q}_s\left[ n \right] \|}{\delta_t}$ 
and 
$V_z\left[ n \right] = \frac{ z_s\left[ n + 1 \right] - z_s\left[ n \right] }{\delta_t}$ denote the horizontal and vertical speeds of $S$ in the $n$-th time slot, respectively, 
$P_{\mathrm{b}}$ and $P_{\mathrm{i}}$ represent the blade profile power and induced power in hover, respectively, 
$U_{\mathrm{tip}}$ is the rotor blade tip speed, $v_0$ denotes the mean rotor-induced velocity in hover, 
$d_0$, $\rho$, $s$, $A$ represent the fuselage drag ratio, air density, rotor solidity, and rotor disc area, respectively, 
and $W$ denotes the weight of $S$.
When $V_{z}\left[ n \right] < 0$, there is $P^{\mathrm{ver}}\left[ n \right] = 0$.

\section{Communication-centric Design}
\label{sec:CommunicationCen}

In this section, we consider communication-centric scenarios, such as post-disaster emergency communications and intelligent transportation management, where the average secrecy rate (ASR) is maximized through the joint optimization of user/target scheduling, computational frequency allocation, transmit beamforming, and UAV trajectory. 
For notational convenience, we define 
$\bm{\theta} = \{ \theta_{U_k}\left[ n \right], \theta_{V_j}\left[ n \right], \forall n, k, j \}$ as scheduling variables for users and sensing targets,	
$\mathbf{F} = \{ f_s\left[ n \right], \forall n \}$ as local computational frequency of $S$,
$\tilde{\mathbf{W}} = \{ \mathbf{W}_i\left[ n \right], \forall i, n \}$ as beamforming matrices for $i \in \{K, J, B\}$,
$\mathbf{Q}_s = \{ \mathbf{q}_s\left[ n \right], \forall n \}$ as horizontal trajectory of $S$, 
and 
$\mathbf{Z} = \{ z_s\left[ n \right], \forall n \}$ as vertical altitude of $S$.

The optimization problem is then formulated as 
\begin{subequations}\label{Opt1.0}
	\begin{align}
		\mathcal{P}_{1.0}: \quad &\max_{\bm{\theta}, \mathbf{F}, \tilde{\mathbf{W}}, \mathbf{Q}_s, \mathbf{Z}} \; {\bar{\mathcal{R}}}^{\mathrm{sec}} \label{Opt10a}\\
		\text{s.t.} \quad & R_{U_k}^{\mathrm{sec}}\left[ n \right] \geq \theta_{U_k}\left[ n \right] R_{\min}^{\mathrm{tra}}, \forall k, n, \label{Opt10b} \\
		& R_{V_j}^{\mathrm{rad}}\left[ n \right] \geq \theta_{V_j}\left[ n \right] R_{\min}^{\mathrm{rad}}, \forall j, n, \label{Opt10c} \\
		& D_s\left[ n \right] \geq D_J\left[ n \right], \forall n, \label{Opt10d} \\
		& f_{\min} \leq f_s\left[ n \right] \leq f_{\max}, \forall n, \label{Opt10e} \\
		& \mathbf{W}_i\left[ n \right] \succeq 0, \forall i, n, \label{Opt10f} \\
		& \operatorname{rank}(\mathbf{W}_i\left[ n \right]) = 1, \forall i, n, \label{Opt10g} \\
		& P^{\mathrm{tra}}\left[ n \right] \leq P_{\max}^{\mathrm{tra}}, \forall n, \label{Opt10h} \\
		& E^{\mathrm{UAV}} \leq E_{\max}^{\mathrm{UAV}}, \label{Opt10i} \\
		& \| \mathbf{q}_s\left[ n + 1 \right] - \mathbf{q}_s\left[ n \right] \| \leq \delta_t V_{\max}^{\mathrm{h}}, \forall n, \label{Opt10j} \\
		& | z_s\left[ n + 1 \right] - z_s\left[ n \right] | \leq \delta_t V_{\max}^{\mathrm{z}}, \forall n, \label{Opt10k} \\
		& z_{\min} \leq z_s\left[ n \right] \leq z_{\max}, \forall n, \label{Opt10l} \\
		& (\mathbf{q}_s\left[ 1 \right], z_s\left[ 1 \right]) = (\mathbf{q}_\mathrm{I}, z_\mathrm{I}), \nonumber\\
		& (\mathbf{q}_s\left[ N \right], z_s\left[ N \right]) = (\mathbf{q}_\mathrm{F}, z_\mathrm{F}), \label{Opt10m} \\
		& (\mathrm{\ref{scheduling1a}})- (\mathrm{\ref{scheduling2b}}), \nonumber
	\end{align}
\end{subequations}
where 
${\bar{\mathcal{R}}}^{\mathrm{sec}} = \frac{1}{N}\sum\limits_{n = 1}^N \sum\limits_{k = 1}^K {R_{{U_k}}^{{\rm{sec}}}}  \left[ n \right]$ denotes the ASR. Constraint (\ref{Opt10b}) ensures that the achievable rate of each scheduled user meets a minimum communication rate requirement $R_{\min}^{\mathrm{tra}}$, while (\ref{Opt10c}) imposes a similar requirement on the radar estimation rate for each sensed target, with threshold $R_{\min}^{\mathrm{rad}}$. 
Constraint (\ref{Opt10d}) guarantees that the amount of data processed locally by $S$ is sufficient to handle the sensed data, where $D_J\left[ n \right] = \sum_{j=1}^J \theta_{V_j}\left[ n \right] D_J$. 
Constraint (\ref{Opt10e}) limits the computational frequency of $S$ within a feasible range. 
Constraints (\ref{Opt10f}) and (\ref{Opt10g}) enforce the positive semidefinite and rank-one conditions on the beamforming matrices, respectively. 
The transmit power and total energy budgets of $S$ are imposed by (\ref{Opt10h}) and (\ref{Opt10i}), 
where $P_{\max}^{\mathrm{tra}}$ and $E_{\max}^{\mathrm{UAV}}$ denote the maximum allowable transmit power and total energy consumption, respectively. 
Constraints (\ref{Opt10j}) and (\ref{Opt10k}) limit the horizontal and vertical speeds of $S$ to $V_{\max}^{\mathrm{h}}$ and $V_{\max}^{\mathrm{z}}$, respectively. 
Constraint (\ref{Opt10l}) restricts the flight altitude within $[z_{\min}, z_{\max}]$, and (\ref{Opt10m}) specifies the initial and final positions of the UAV. 
Finally, constraints (\ref{scheduling1a}) - (\ref{scheduling2b}) govern the scheduling of users and targets.

It is worth noting that $\mathcal{P}_{1.0}$ is a highly non-convex optimization problem due to the coupling among scheduling variables, beamforming matrices, trajectory parameters, and computational frequency, making it mathematically intractable to solve directly.
To tackle this challenge, we employ an alternating optimization (AO) approach, which iteratively optimizes one block of variables, namely $\bm{\theta}$, $\mathbf{F}$, $\tilde{\mathbf{W}}$, $\mathbf{Q}_s$, and $\mathbf{Z}$, while keeping the others fixed.
For notational simplicity, we introduce the following auxiliary variables:
$\mathbf{A}_l\left[ n \right] = \mathbf{a}_l\left[ n \right] \mathbf{a}_l^H\left[ n \right]$, 
$S_{U_k}\left[ n \right] = \operatorname{tr}\left( \mathbf{A}_{U_k}\left[ n \right] \mathbf{W}_K\left[ n \right] \right)$, 
$S_E\left[ n \right] = \operatorname{tr}\left( \mathbf{A}_E\left[ n \right] \mathbf{W}_K\left[ n \right] \right)$, 
$S_B\left[ n \right] = \operatorname{tr}\left( \mathbf{A}_B\left[ n \right] \mathbf{W}_B\left[ n \right] \right)$, 
$I_{U_k}\left[ n \right] = \operatorname{tr}\left( \mathbf{A}_{U_k}\left[ n \right] \left( \mathbf{W}_J\left[ n \right] + \mathbf{W}_B\left[ n \right] \right) \right)$, 
$I_E\left[ n \right] = \operatorname{tr}\left( \mathbf{A}_E\left[ n \right] \left( \mathbf{W}_J\left[ n \right] + \mathbf{W}_B\left[ n \right] \right) \right)$, 
$I_B\left[ n \right] = \operatorname{tr}\left( \mathbf{A}_B\left[ n \right] \left( \mathbf{W}_K\left[ n \right] + \mathbf{W}_J\left[ n \right] \right) \right)$.

Based on \cite{LeiH2025IoTCongke}, $R_{U_k}^{\mathrm{sec}}\left[ n \right]$ is replacing as
\begin{align}
	{\hat R}_{U_k}^{\mathrm{sec}}\left[ n \right] & =  \theta_{U_k}\left[ n \right] P_{U_k}^{\mathrm{L}}\left[ n \right] \log_2 \left( 1 + \frac{S_{U_k}\left[ n \right]}{I_{U_k}\left[ n \right] + \mu_{U_k}\left[ n \right]} \right) \nonumber \\
	&- \theta_{U_k}\left[ n \right] P_{E}^{\mathrm{L}}\left[ n \right] \log_2 \left( 1 + \frac{S_E\left[ n \right]}{I_E\left[ n \right] + \mu_E\left[ n \right]} \right).
\end{align}

\subsection{Communication and Sensing Scheduling Optimization}

In this subsection, we optimize the scheduling variables $\bm{\theta}$ with $\{\mathbf{F}, \tilde{\mathbf{W}}, \mathbf{Q}_s, \mathbf{Z}\}$ fixed. 
To facilitate the optimization, the original binary constraints 
$\theta_{U_k}\left[ n \right] \in \{0, 1\}$ 
and 
$\theta_{V_j}\left[ n \right] \in \{0, 1\}$ 
are relaxed to 
$0 \leq \theta_{U_k}\left[ n \right] \leq 1$ 
and 
$0 \leq \theta_{V_j}\left[ n \right] \leq 1$, respectively. 
Accordingly, $\mathcal{P}_{1.0}$ is reformulated as
\begin{subequations}
	\begin{align}
		\mathcal{P}_{1.1{\rm a}}: \quad &\max_{\bm{\theta}} \; \frac{1}{N}\sum\limits_{n = 1}^N \sum\limits_{k = 1}^K {\hat R}_{U_k}^{\mathrm{sec}}\left[ n \right]  \label{P1.1a} \\
		\text{s.t.} \quad &{\hat R}_{U_k}^{\mathrm{sec}}\left[ n \right] \geq \theta_{U_k}\left[ n \right] R_{\min}^{\mathrm{tra}}, \forall k, n, \label{P1.1b} \\
		&0 \leq \theta_{U_k}\left[ n \right] \leq 1, \label{scheduling1b2}\\
		&0 \leq \theta_{V_j}\left[ n \right] \leq 1, \label{scheduling2b2}\\
		& \mathrm{(\ref{Opt10c}), (\ref{Opt10d}),} \mathrm{(\ref{scheduling1a}), (\ref{scheduling2a})}. \nonumber
	\end{align}
\end{subequations}
In $\mathcal{P}_{1.1{\rm a}}$, constraint (\ref{Opt10c}) is non-convex with respect to ${\bm{\theta}}$, which makes $\mathcal{P}_{1.1{\rm a}}$ a non-convex optimization problem.

Constraint (\ref{Opt10c}) is equivalently rewritten as
\begin{align}
	R_{V_j}^{\mathrm{rad}}\left[ n \right] = \frac{\theta_{V_j}\left[ n \right]}{2\mu \delta^{-1}} \log_2 \left( 1 + \frac{\Gamma_{V_j}^{\mathrm{rad}}\left[ n \right]}{d_{V_j}^4\left[ n \right]} \right) \geq \theta_{V_j}\left[ n \right] \frac{R_{\min}^{\mathrm{rad}}}{P_{V_j}^{\mathrm{L}}\left[ n \right]},
	\label{P1.1C2}
\end{align}
where 
$\Gamma_{V_j}^{\mathrm{rad}}\left[ n \right] = \frac{2 \sigma_{\mathrm{pre}}^2 \hat{\gamma}^2 B_w^3 \mu \xi}{\beta_0^{-1} \sigma_s^2} \operatorname{tr}\left( \mathbf{A}_{V_j}\left[ n \right] \mathbf{W}_J\left[ n \right] \mathbf{A}_{V_j}^H\left[ n \right] \right).$
Consequently, $\mathcal{P}_{1.1}$ is reformulated as
\begin{subequations}
	\begin{align}
		\mathcal{P}_{1.1{\rm b}}: \quad &\max_{\bm{\theta}} \; \frac{1}{N}\sum\limits_{n = 1}^N \sum\limits_{k = 1}^K {\hat R}_{U_k}^{\mathrm{sec}}\left[ n \right]  \\
		\text{s.t.} \quad & \mathrm{\mathrm{(\ref{scheduling1a}), (\ref{scheduling2a})}, (\ref{Opt10d}), \mathrm{(\ref{P1.1a})}-\mathrm{(\ref{scheduling2b2}), (\ref{P1.1C2})}}. \nonumber
	\end{align}
\end{subequations}
$\mathcal{P}_{1.1{\rm b}}$ is a linear problem and can be efficiently solved using standard optimization tools such as CVX.

\subsection{Computational Frequency and Power Allocation Optimization }

In this subsection, we optimize the computational frequency $\mathbf{F}$ and the beamforming matrices $\tilde{\mathbf{W}}$ with $\{\bm{\theta}, \mathbf{Q}_s, \mathbf{Z}\}$ fixed. The corresponding optimization problem is formulated as
\begin{subequations}
\begin{align}
	\mathcal{P}_{1.2{\rm a}}: \quad &\max_{\mathbf{F}, \tilde{\mathbf{W}}} \; \frac{1}{N}\sum\limits_{n = 1}^N \sum\limits_{k = 1}^K 
	{\hat R}_{U_k}^{\mathrm{sec}}\left[ n \right] \label{P1.2a} \\
	\text{s.t.} \quad & \mathrm{ (\ref{Opt10d})-(\ref{Opt10i}), (\ref{P1.1b})}, \mathrm{(\ref{P1.1C2})}.
\end{align}
\end{subequations}
In $\mathcal{P}_{1.2{\rm a}}$, the objective function (\ref{P1.2a}) along with constraints (\ref{Opt10d}), (\ref{Opt10g}), and (\ref{P1.1b}), are all non-convex with respect to $\tilde{\mathbf{W}}$, rendering $\mathcal{P}_{1.2{\rm a}}$ a non-convex optimization problem. 

To deal with ${\hat R}_{U_k}^{\mathrm{sec}}\left[ n \right]$ in (\ref{P1.1b}) and (\ref{P1.2a}), we apply a first-order Taylor expansion to obtain a concave lower bound ${\hat R}_{U_k}^{\mathrm{sec,lb}}\left[ n \right]$, given by
\begin{align}
	{\hat R}_{U_k}^{\mathrm{sec,lb}}\left[ n \right] & =  \theta_{U_k}\left[ n \right] P_{U_k}^{\mathrm{L}}\left[ n \right] \log_2 \left( S_{U_k}\left[ n \right] + I_{U_k}\left[ n \right] + \mu_{U_k}\left[ n \right] \right) \nonumber \\
	&- \theta_{U_k}\left[ n \right] \left( P_{U_k}^{\mathrm{L}}\left[ n \right] R_{U_k}^{\mathrm{sec,a}}\left[ n \right] + P_E^{\mathrm{L}}\left[ n \right] R_E^{\mathrm{sec,b}}\left[ n \right] \right) \nonumber \\
	&+ \theta_{U_k}\left[ n \right] P_E^{\mathrm{L}}\left[ n \right] \log_2 \left( I_E\left[ n \right] + \mu_E\left[ n \right] \right),
\end{align}
where
$R_{U_k}^{\mathrm{sec,a}}\left[ n \right] = \log_2 \left( I_{U_k}^{\left( m \right)}\left[ n \right] + \mu_{U_k}\left[ n \right] \right) + \frac{I_{U_k}\left[ n \right] - I_{U_k}^{\left( m \right)}\left[ n \right]}{\ln 2 \left( I_{U_k}^{\left( m \right)}\left[ n \right] + \mu_{U_k}\left[ n \right] \right)}$, 
$R_E^{\mathrm{sec,b}}\left[ n \right] = \log_2 \left( S_E^{\left( m \right)}\left[ n \right] + I_E^{\left( m \right)}\left[ n \right] + \mu_E\left[ n \right] \right)
	 + \frac{\left( S_E\left[ n \right] - S_E^{\left( m \right)}\left[ n \right] \right) + \left( I_E\left[ n \right] - I_E^{\left( m \right)}\left[ n \right] \right)}{\ln 2 \left( S_E^{\left( m \right)}\left[ n \right] + I_E^{\left( m \right)}\left[ n \right] + \mu_E\left[ n \right] \right)}$, 
and 
$X^{\left( m \right)}$ denotes the feasible point of $X$ at the $m$-th iteration.

The constraint (\ref{Opt10d}) is rewritten as
\begin{align}
	R_B^{\mathrm{app}}\left[ n \right] &= {P_B^{\mathrm{L}}\left[ n \right]}\log_2 \left( S_B\left[ n \right] + I_B\left[ n \right] + \mu_B\left[ n \right] \right) \nonumber \\
	&- {P_B^{\mathrm{L}}\left[ n \right]}\log_2 \left( I_B\left[ n \right] + \mu_B\left[ n \right] \right) \geq {\Gamma^f\left[ n \right]},
\end{align}
where $\Gamma^f\left[ n \right] = \frac{D_J\left[ n \right]}{B_w} - \frac{f_s\left[ n \right]}{F_s B_w}$. Since the second term $\log_2(I_B\left[ n \right] + \mu_B\left[ n \right])$ is non-concave, we replace it with its first-order Taylor expansion to obtain a convex approximation:
\begin{align}
	\tilde{R}_B^{\mathrm{app}}\left[ n \right] & =  {P_B^{\mathrm{L}}\left[ n \right]}\log_2 \left( S_B\left[ n \right] + I_B\left[ n \right] + \mu_B\left[ n \right] \right) \nonumber \\
	&- {P_B^{\mathrm{L}}\left[ n \right]}\log_2 \left( I_B^{\left( m \right)}\left[ n \right] + \mu_B\left[ n \right] \right) \nonumber \\
	&- {P_B^{\mathrm{L}}\left[ n \right]}\frac{I_B\left[ n \right] - I_B^{\left( m \right)}\left[ n \right]}{\ln 2 \left( I_B^{\left( m \right)}\left[ n \right] + \mu_B\left[ n \right] \right)} \geq {\Gamma^f\left[ n \right]}.
\end{align}
By dropping the rank-one constraint in (\ref{Opt10g}), $\mathcal{P}_{1.2}$ is reformulated as
\begin{subequations}
	\begin{align}
		\mathcal{P}_{1.2{\rm b}}: \quad &\max_{\mathbf{F}, \tilde{\mathbf{W}}} \; \frac{1}{N}\sum\limits_{n = 1}^N \sum\limits_{k = 1}^K {\hat R}_{U_k}^{\mathrm{sec,lb}}\left[ n \right] \\
		\text{s.t.} \quad & {\hat R}_{U_k}^{\mathrm{sec,lb}}\left[ n \right] \geq \theta_{U_k}\left[ n \right] R_{\min}^{\mathrm{tra}}, \forall k, n, \label{P1.2C1} \\
		& \tilde{R}_B^{\mathrm{app}}\left[ n \right] \geq {\Gamma^f\left[ n \right]}, \forall n, \label{P1.2C3} \\
		& \mathrm{(\ref{Opt10e}), (\ref{Opt10f}), (\ref{Opt10h}), (\ref{Opt10i}), (\ref{P1.1C2})}. \nonumber
	\end{align}
\end{subequations}
$\mathcal{P}_{1.2{\rm b}}$ is now a semidefinite programming (SDP) problem and can be efficiently solved using CVX. To recover a rank-one solution, standard techniques such as Gaussian randomization, singular value decomposition, or the method proposed in \cite{DanQ2026TVT} can be applied.

\subsection{Horizontal Trajectory Optimization of $S$} 

In this subsection, we optimize the horizontal trajectory $\mathbf{Q}_s$ with $\{\bm{\theta}, \mathbf{F}, \tilde{\mathbf{W}}, \mathbf{Z}\}$ fixed. Accordingly, $\mathcal{P}_{1.3}$ is reformulated as
\begin{subequations}
	\begin{align}
		\mathcal{P}_{1.3{\rm a}}: \quad &\max_{\mathbf{Q}_s} \; \frac{1}{N}\sum\limits_{n = 1}^N \sum\limits_{k = 1}^K {\hat R}_{U_k}^{\mathrm{sec}}\left[ n \right] \label{P1.3a}\\
		\text{s.t.} \quad & \mathrm{(\ref{Opt10c})-(\ref{Opt10d}), (\ref{P1.1b}), (\ref{Opt10i}), (\ref{Opt10j}), (\ref{Opt10m})}.
	\end{align}
\end{subequations}
In $\mathcal{P}_{1.3{\rm a}}$, the objective function (\ref{P1.3a}) along with constraints (\ref{Opt10c})-(\ref{Opt10d}), (\ref{Opt10i}), and (\ref{P1.1b}) are all non-convex with respect to $\mathbf{Q}_s$, rendering $\mathcal{P}_{1.3{\rm a}}$ a non-convex optimization problem.

It must be noted that $\mathbf{A}_l\left[ n \right]$ is a complicated nonlinear function of the UAV trajectory variables, making the trajectory design highly challenging. To make the problem more tractable, we approximate $\mathbf{A}_l\left[ n \right]$ in the $(m+1)$-th iteration using the trajectory from the $m$-th iteration \cite{LeiH2025IoTCongke, DengC2023TWC}. 
To decouple the strong coupling between $\mathbf{Q}_s$ and $P_l^{\mathrm{L}}\left[ n \right]$, the following constraints are introduced \cite{LeiH2024TCCN3D}:
\begin{subequations}
	\begin{align}
		&1 + C e^{CD} e^{-D \phi_l^{\mathrm{h}}\left[ n \right]} \leq p_l^{\mathrm{h}}\left[ n \right], \forall l \in \{\mathcal{L} \setminus E\}, \label{C1a1h} \\
		&p_E^{\mathrm{h}}\left[ n \right] \leq 1 + C e^{CD} e^{-D \phi_E^{\mathrm{h}}\left[ n \right]}, \label{C1a2hm} \\
		&\phi_l^{\mathrm{h}}\left[ n \right] \leq \frac{180}{\pi} \arctan\left( \frac{z_s\left[ n \right]}{\varphi_l^{\mathrm{h},1}\left[ n \right]} \right), \forall l \in \{\mathcal{L} \setminus E\}, \label{C1a3hm} \\
		&\frac{180}{\pi} \arctan\left( \frac{z_s\left[ n \right]}{\varphi_E^{\mathrm{h},1}\left[ n \right]} \right) \leq \phi_E^{\mathrm{h}}\left[ n \right], \label{C1a4h} \\
		&\| \mathbf{q}_s\left[ n \right] - \mathbf{q}_l \| \leq \varphi_l^{\mathrm{h},1}\left[ n \right], \forall l \in \{\mathcal{L} \setminus E\}, \label{C1a5h} \\
		&\varphi_E^{\mathrm{h},1}\left[ n \right] \leq \| \mathbf{q}_s\left[ n \right] - \mathbf{q}_E \|,  \label{C1a6hm}
	\end{align}
\end{subequations}
where $\{ p_l^{\mathrm{h}}\left[ n \right], \phi_l^{\mathrm{h}}\left[ n \right], \varphi_l^{\mathrm{h},1}\left[ n \right], \forall l, n \}$ are newly introduced slack variables. 
Since $e^{-D \phi_E^{\mathrm{h}}\left[ n \right]}$, $\arctan\left( \frac{z_s\left[ n \right]}{\varphi_l^{\mathrm{h},1}\left[ n \right]} \right)$, and $\| \mathbf{q}_s\left[ n \right] - \mathbf{q}_E \|$ are convex, constraints (\ref{C1a2hm}), (\ref{C1a3hm}), and (\ref{C1a6hm}) are non-convex with respect to $\mathbf{Q}_s$. 
To address this, we introduce auxiliary variables $F_l^{\mathrm{h}}\left[ n \right]$ and apply the SCA technique to transform them into the following linear forms, 
\begin{subequations}
	\begin{align}
		p_E^{\mathrm{h}}\left[ n \right] &\leq 1 + C e^{CD} F_E^{\mathrm{h}}\left[ n \right], \label{C1a2h} \\
		\phi_l^{\mathrm{h}}\left[ n \right] &\leq \frac{180}{\pi} F_l^{\mathrm{h}}\left[ n \right], \forall l \in \{\mathcal{L} \setminus E\}, \label{C1a3h} \\
		\varphi_E^{\mathrm{h},1}\left[ n \right] &\leq \| {\mathbf{q}}_s^{\left( m \right)}\left[ n \right] - \mathbf{q}_E \| \nonumber \\
		&+ \frac{ \left( {\mathbf{q}}_s^{\left( m \right)}\left[ n \right] - \mathbf{q}_E \right)^T \left( {\mathbf{q}}_s \left[ n \right] - {\mathbf{q}}_s^{\left( m \right)}\left[ n \right] \right) }{ \| {\mathbf{q}}_s^{\left( m \right)}\left[ n \right] - \mathbf{q}_E \| } , \label{C1a6h}
	\end{align}
\end{subequations}
with
\begin{subequations}
	\begin{align}
		F_l^{\mathrm{h}}\left[ n \right] &= \arctan\left( \frac{z_s\left[ n \right]}{\varphi_l^{\mathrm{h},1,\left( m \right)}\left[ n \right]} \right) \nonumber \\
		&- \frac{z_s\left[ n \right] \left( \varphi_l^{\mathrm{h},1}\left[ n \right] - \varphi_l^{\mathrm{h},1,\left( m \right)}\left[ n \right] \right)}{ \left( \varphi_l^{\mathrm{h},1,\left( m \right)}\left[ n \right] \right)^2 + z_s^2\left[ n \right] }, \forall l \in \{\mathcal{L} \setminus E\}, \\	
		F_E^{\mathrm{h}}\left[ n \right] &= e^{-D \phi_E^{\mathrm{h},\left( m \right)}\left[ n \right]} \left( 1 - D \left( \phi_E^{\mathrm{h}}\left[ n \right] - \phi_E^{\mathrm{h},\left( m \right)}\left[ n \right] \right) \right).
		\label{H232}
	\end{align}
\end{subequations}

Similarly, since $d_l^2\left[ n \right] = z_s^2\left[ n \right] + d_{h,l}^2\left[ n \right]$ is convex, it introduces non-convexity into the objective and constraints. For legitimate users, we introduce auxiliary variables $\varphi_{U_k}^{\mathrm{h},2}\left[ n \right]$ and $\varphi_E^{\mathrm{h},2}\left[ n \right]$ satisfying
\begin{subequations}
\begin{align}
	&\varphi_{U_k}^{\mathrm{h},2}\left[ n \right] \geq z_s^2\left[ n \right] + d_{h,U_k}^2\left[ n \right], \label{C1a7h}\\
	&\varphi_E^{\mathrm{h},2}\left[ n \right] \leq \left( d_{h,E}^2\left[ n \right] \right)^{\mathrm{lb}} + z_s^2\left[ n \right], \label{C1a8h}
\end{align}
\end{subequations}
where $\left( d_{h,E}^2\left[ n \right] \right)^{\mathrm{lb}} = 2 \left( \mathbf{q}_s^{\left( m \right)}\left[ n \right] - \mathbf{q}_E \right)^T $ $ \left( \mathbf{q}_s\left[ n \right] - \mathbf{q}_s^{\left( m \right)}\left[ n \right] \right) + \| \mathbf{q}_s^{\left( m \right)}\left[ n \right] - \mathbf{q}_E \|^2 $.

Firstly, ${\hat R}_{U_k}^{\mathrm{sec}}\left[ n \right]$ in (\ref{P1.1b}) and (\ref{P1.3a}) is decomposed as
\begin{align}
	{\tilde R}_{U_k}^{\mathrm{sec}}\left[ n \right] = G_{U_k}^{\mathrm{h}}\left[ n \right] - G_E^{\mathrm{h}}\left[ n \right],
	\label{H232}
\end{align} 
where
\begin{subequations}
\begin{align}
	G_{U_k}^{\mathrm{h}}\left[ n \right] &= \frac{1}{p_{U_k}^{\mathrm{h}}\left[ n \right]} \log_2 \left( 1 + \frac{ \theta_{U_k}\left[ n \right] S_{U_k}\left[ n \right] }{ I_{U_k}\left[ n \right] + \mu_{U_k}^{\mathrm{h}}\left[ n \right] } \right), \\
	G_E^{\mathrm{h}}\left[ n \right] &= \frac{1}{p_E^{\mathrm{h}}\left[ n \right]} \log_2 \left( 1 + \frac{ \theta_{U_k}\left[ n \right] S_E\left[ n \right] }{ I_E\left[ n \right] + \mu_E^{\mathrm{h}}\left[ n \right] } \right),
	\label{H232}
\end{align}
\end{subequations}
with $\mu_{U_k}^{\mathrm{h}}\left[ n \right] = \beta_0^{-1} \sigma_{U_k}^2 \varphi_{U_k}^{\mathrm{h},2}\left[ n \right]$ and $\mu_E^{\mathrm{h}}\left[ n \right] = \beta_0^{-1} \sigma_E^2 \varphi_E^{\mathrm{h},2}\left[ n \right]$. 

According to \cite{LeiH2024TCCN3D}, for any $A \geq 0$, $f(x, y) = \frac{1}{x} \log_2 \left( 1 + \frac{A}{y} \right)$ is jointly convex in $x > 0$ and $y > 0$. Thus, by applying a first-order Taylor expansion, we obtain a concave lower bound for ${\tilde R}_{U_k}^{\mathrm{sec}}\left[ n \right]$
\begin{align}
	\tilde{R}_{U_k}^{\mathrm{sec,lb,h}}\left[ n \right] & =  G_{U_k,1}^{\mathrm{h},\left( m \right)}\left[ n \right] + G_{U_k,2}^{\mathrm{h},\left( m \right)}\left[ n \right] \left( p_{U_k}^{\mathrm{h}}\left[ n \right] - p_{U_k}^{\mathrm{h},\left( m \right)}\left[ n \right] \right) \nonumber \\
	&+ G_{U_k,3}^{\mathrm{h},\left( m \right)}\left[ n \right] \left( \mu_{U_k}^{\mathrm{h}}\left[ n \right] - \mu_{U_k}^{\mathrm{h},\left( m \right)}\left[ n \right] \right) - G_E^{\mathrm{h}}\left[ n \right],
\end{align}
where 
\begin{subequations}
\begin{align}
	G_{U_k,1}^{\mathrm{h},\left( m \right)}\left[ n \right] &= \frac{ \log_2 \left( 1 + \gamma_{U_k}^{\mathrm{h},\left( m \right)}\left[ n \right] \right) }{ p_{U_k}^{\mathrm{h},\left( m \right)}\left[ n \right] }, \\
	G_{U_k,2}^{\mathrm{h},\left( m \right)}\left[ n \right] &= -\frac{ \log_2 \left( 1 + \gamma_{U_k}^{\mathrm{h},\left( m \right)}\left[ n \right] \right) }{ \left( p_{U_k}^{\mathrm{h},\left( m \right)}\left[ n \right] \right)^2 }, \\
	G_{U_k,3}^{\mathrm{h},\left( m \right)}\left[ n \right] &= \frac{ - \theta_{U_k}\left[ n \right] S_{U_k}\left[ n \right] / \left( 1 + \gamma_{U_k}^{\mathrm{h},\left( m \right)}\left[ n \right] \right) }{ \ln 2 \cdot p_{U_k}^{\mathrm{h},\left( m \right)}\left[ n \right] \left( I_{U_k}\left[ n \right] + \mu_{U_k}^{\mathrm{h},\left( m \right)}\left[ n \right] \right)^2 }, \\
	\gamma_{U_k}^{\mathrm{h},\left( m \right)}\left[ n \right] &= \frac{ \theta_{U_k}\left[ n \right] S_{U_k}\left[ n \right] }{ I_{U_k}\left[ n \right] + \mu_{U_k}^{\mathrm{h},\left( m \right)}\left[ n \right] }.
\end{align}
\end{subequations}

Then, (\ref{Opt10c}) is rewritten as
\begin{align}
	2 \theta_{V_j}\left[ n \right] \mu \delta^{-1} R_{\min}^{\mathrm{rad}} p_{V_j}^{\mathrm{h}}\left[ n \right] \leq \log_2 \left( 1 + \frac{ \Gamma_{V_j}^{\mathrm{rad}}\left[ n \right] }{ d_{V_j}^4\left[ n \right] } \right). \label{C2rqm}
\end{align}
It should be noted that the right-hand side of (\ref{C2rqm}) is convex, making it a non-convex constraint. By introducing the slack variable $\varphi_{V_j}^{\mathrm{h},2}\left[ n \right]$ and 
applying SCA, we obtain the convex approximation:
\begin{align}
	\log_2 \left( 1 + \frac{ \Gamma_{V_j}^{\mathrm{rad}}\left[ n \right] }{ \varphi_{V_j}^{\mathrm{h},2,\left( m \right)}\left[ n \right] } \right) &- \frac{ \Gamma_{V_j}^{\mathrm{rad}}\left[ n \right] \left( \frac{ \varphi_{V_j}^{\mathrm{h},2}\left[ n \right] }{ \varphi_{V_j}^{\mathrm{h},2,\left( m \right)}\left[ n \right] } - 1 \right) }{ \ln 2 \left( \Gamma_{V_j}^{\mathrm{rad}}\left[ n \right] + \varphi_{V_j}^{\mathrm{h},2,\left( m \right)}\left[ n \right] \right) } \nonumber \\
	& \geq 2 \theta_{V_j}\left[ n \right] \mu \delta^{-1} R_{\min}^{\mathrm{rad}} p_{V_j}^{\mathrm{h}}\left[ n \right], \label{C2rq}
\end{align}
with
\begin{align}
	z_s^2\left[ n \right] + \| \mathbf{q}_s\left[ n \right] - \mathbf{q}_{V_j} \|^2 \leq \sqrt{ \varphi_{V_j}^{\mathrm{h},2}\left[ n \right] }. \label{C2a1h}
\end{align}

By introducing the slack variable $\varphi_B^{\mathrm{h},2}\left[ n \right]$, (\ref{Opt10d}) is recast as
\begin{align}
	\Gamma^f\left[ n \right] p_B^{\mathrm{h}}\left[ n \right] \leq \tilde{R}_B^{\mathrm{app,h}}\left[ n \right], \label{C3rq}
\end{align}
where 
$\tilde{R}_B^{\mathrm{app,h}}\left[ n \right] = \log_2 \left( S_B\left[ n \right] + I_B\left[ n \right] + \mu_B^{\mathrm{h}}\left[ n \right] \right) - \log_2 \left( I_B\left[ n \right] + \mu_B^{\mathrm{h},\left( m \right)}\left[ n \right] \right) - \frac{ \mu_B^{\mathrm{h}}\left[ n \right] - \mu_B^{\mathrm{h},\left( m \right)}\left[ n \right] }{ \ln 2 \left( I_B\left[ n \right] + \mu_B^{\mathrm{h},\left( m \right)}\left[ n \right] \right) }$, 
and 
$\mu_B^{\mathrm{h}}\left[ n \right] = \beta_0^{-1} \sigma_B^2 \varphi_B^{\mathrm{h},2}\left[ n \right]$, and 
with the following constraint
\begin{align}
	z_s^2\left[ n \right] + d_{h,B}^2\left[ n \right] \leq \varphi_B^{\mathrm{h},2}\left[ n \right]. \label{C3a1h}
\end{align}

Following \cite{ZengY2019TWC}, the energy constraint (\ref{Opt10i}) is rewritten as
\begin{align}
	\sum_{n=1}^{N} \delta_t \left( \tilde{P}^{\mathrm{fly}}\left[ n \right] + \kappa f_s^3\left[ n \right] + P^{\mathrm{tra}}\left[ n \right] \right) \leq E_{\max}^{\mathrm{UAV}}, \label{P1.3C8}
\end{align}
where 
$\tilde{P}^{\mathrm{fly}}\left[ n \right] = P_{\mathrm{i}} v_2\left[ n \right] + \frac{1}{2} d_0 \rho s A v_1^3\left[ n \right] + P_{\mathrm{b}} \left( 1 + \frac{3 v_1^2\left[ n \right]}{U_{\mathrm{tip}}^2} \right) + W V_z\left[ n \right]$, 
$v_1\left[ n \right]$ and $v_2\left[ n \right]$ are the auxiliary variables satisfying
\begin{subequations}
\begin{align}
	&v_1\left[ n \right] \geq \frac{ \| \mathbf{q}_s\left[ n + 1 \right] - \mathbf{q}_s\left[ n \right] \| }{ \delta_t }, \label{v1} \\
	&v_2^2\left[ n \right] + \frac{ \| \mathbf{q}_s\left[ n + 1 \right] - \mathbf{q}_s\left[ n \right] \|^2 }{ v_0^2 \delta_t^2 } \geq \frac{1}{ v_2^2\left[ n \right] }. \label{v2}
\end{align}
\end{subequations}
Since (\ref{v2}) is non-convex due to the convex left-hand side, we apply SCA to obtain the convex approximation (\ref{v2t}), shown at the top of the next page.

\begin{figure*}[ht]
	\begin{align}
		&\left( v_2^{\left( m \right)}\left[ n \right] \right)^2 + 2 v_2^{\left( m \right)}\left[ n \right] \left( v_2\left[ n \right] - v_2^{\left( m \right)}\left[ n \right] \right) + \frac{ \| \mathbf{q}_s^{\left( m \right)}\left[ n + 1 \right] - \mathbf{q}_s^{\left( m \right)}\left[ n \right] \|^2 }{ v_0^2 \delta_t^2 } \nonumber \\
		&+ \frac{2}{v_0^2 \delta_t^2} \left( \mathbf{q}_s^{\left( m \right)}\left[ n + 1 \right] - \mathbf{q}_s^{\left( m \right)}\left[ n \right] \right)^T \left( \mathbf{q}_s\left[ n + 1 \right] - \mathbf{q}_s\left[ n \right] - \mathbf{q}_s^{\left( m \right)}\left[ n + 1 \right] + \mathbf{q}_s^{\left( m \right)}\left[ n \right] \right) \geq \frac{1}{ v_2^2\left[ n \right] } \label{v2t}
	\end{align}
	\hrulefill
\end{figure*}

Finally, $\mathcal{P}_{1.3{\rm a}}$ is reformulated as the following convex problem:
\begin{subequations}
\begin{align}
	\mathcal{P}_{1.3{\rm b}}: \quad &\max_{\mathbf{Q}_s, \mathbf{p}^{\mathrm{h}}, \bm{\varphi}^{\mathrm{h}}, \mathbf{v}} \; \frac{1}{N}\sum\limits_{n = 1}^N \sum\limits_{k = 1}^K \tilde{R}_{U_k}^{\mathrm{sec,lb,h}}\left[ n \right] \\ 
	\text{s.t.} \quad & \tilde{R}_{U_k}^{\mathrm{sec,lb,h}}\left[ n \right] \geq \theta_{U_k}\left[ n \right] R_{\min}^{\mathrm{tra}}, \forall k, n, \label{C1rq} \\
	& \mathrm{(\ref{Opt10j}), (\ref{Opt10m}), (\ref{C1a1h}), (\ref{C1a4h}), (\ref{C1a5h})}, \nonumber\\
	& \mathrm{(\ref{C1a2h})-(\ref{C1a6h}), (\ref{C1a7h}), (\ref{C1a8h})}, \nonumber \\
	& \mathrm{(\ref{C2rq})-(\ref{v1}), (\ref{v2t})}. \nonumber
\end{align}
\end{subequations}
where $\mathbf{p}^{\mathrm{h}} = \{ p_l^{\mathrm{h}}\left[ n \right], \phi_l^{\mathrm{h}}\left[ n \right], \forall l, n \}$, $\bm{\varphi}^{\mathrm{h}} = \{ \varphi_l^{\mathrm{h},1}\left[ n \right], \varphi_l^{\mathrm{h},2}\left[ n \right], \forall l, n \}$, and $\mathbf{v} = \{ v_1\left[ n \right], v_2\left[ n \right], \forall n \}$ are newly introduced slack variable sets. 
$\mathcal{P}_{1.3{\rm b}}$ is convex and can be efficiently solved using CVX.

\subsection{Vertical Trajectory Optimization of $S$} 

In this subsection, we optimize the vertical trajectory $\mathbf{Z}$ with $\{\bm{\theta}, \mathbf{F}, \tilde{\mathbf{W}}, \mathbf{Q}_s\}$ fixed. The optimization problem is formulated as
\begin{subequations}
\begin{align}
	\mathcal{P}_{1.4{\rm a}}: \quad &\max_{\mathbf{Z}} \; \frac{1}{N}\sum\limits_{n = 1}^N \sum\limits_{k = 1}^K {\hat R}_{U_k}^{\mathrm{sec}}\left[ n \right] \label{P1.4a} \\
	\text{s.t.} \quad & \mathrm{(\ref{Opt10c})-(\ref{Opt10d}), (\ref{Opt10i}), (\ref{Opt10k})-(\ref{Opt10m}), (\ref{P1.1b})}. \nonumber
\end{align}
\end{subequations}
In $\mathcal{P}_{1.4{\rm a}}$, the objective function (\ref{P1.4a}) along with constraints (\ref{Opt10c})-(\ref{Opt10d}) and (\ref{P1.1b})  are all non-convex with respect to $\mathbf{Z}$, rendering $\mathcal{P}_{1.4{\rm a}}$ a non-convex optimization problem.

To handle the non-convexity of ${\hat R}_{U_k}^{\mathrm{sec}}\left[ n \right] $ in (\ref{P1.4a}) and (\ref{P1.1b}) introduced by the vertical trajectory, we introduce new slack variable sets $\mathbf{p}^{\mathrm{v}} = \{ p_l^{\mathrm{v}}\left[ n \right], \phi_l^{\mathrm{v}}\left[ n \right], \forall l, n \}$ and $\bm{\varphi}^{\mathrm{v}} = \{ \varphi_l^{\mathrm{v}}\left[ n \right], \forall l, n \}$, which satisfy the following constraints:
\begin{subequations}
	\begin{align}
		&1 + C e^{CD} e^{-D \phi_l^{\mathrm{v}}\left[ n \right]} \leq p_l^{\mathrm{v}}\left[ n \right], \forall l \in \{\mathcal{L} \setminus E\}, \label{C1a1v} \\
		&p_E^{\mathrm{v}}\left[ n \right] \leq 1 + C e^{CD} F_p^{\mathrm{v}}\left[ n \right], \label{C1a2v} \\
		&\phi_l^{\mathrm{v}}\left[ n \right] \leq \frac{180}{\pi} \arctan\left( \frac{z_s\left[ n \right]}{d_{h,l}\left[ n \right]} \right), \forall l \in \{\mathcal{L} \setminus E\}, \label{C1a3v} \\
		&\frac{180}{\pi} F_{\phi}^{\mathrm{v}}\left[ n \right] \leq \phi_E^{\mathrm{v}}\left[ n \right], \label{C1a4v} \\
		&z_s^2\left[ n \right] + d_{h,U_k}^2\left[ n \right] \leq \varphi_{U_k}^{\mathrm{v}}\left[ n \right], \label{C1a5v} \\
		&z_s^2\left[ n \right] + d_{h,B}^2\left[ n \right] \leq \varphi_B^{\mathrm{v}}\left[ n \right], \label{C1a6v} \\
		&\left( z_s^2\left[ n \right] + d_{h,V_j}^2\left[ n \right] \right)^2 \leq \varphi_{V_j}^{\mathrm{v}}\left[ n \right], \label{C2a1v} \\
		&\varphi_E^{\mathrm{v}}\left[ n \right] \leq 2 z_s^{\left( m \right)}\left[ n \right] \left( z_s\left[ n \right] - z_s^{\left( m \right)}\left[ n \right] \right) + d_{h,E}^2\left[ n \right], \label{C3a1v}
	\end{align}
\end{subequations}
where
\begin{subequations}
\begin{align}
	F_p^{\mathrm{v}}\left[ n \right] &= \arctan\left( \frac{z_s\left[ n \right]}{d_E\left[ n \right]} \right) + \frac{ d_E\left[ n \right] \left( z_s\left[ n \right] - z_s^{\left( m \right)}\left[ n \right] \right) }{ d_E^2\left[ n \right] + \left( z_s^{\left( m \right)}\left[ n \right] \right)^2 }, 	\label{FpV} \\
	F_{\phi}^{\mathrm{v}}\left[ n \right] &= e^{-D \phi_E^{\mathrm{v},\left( m \right)}\left[ n \right]} \left( 1 - D \left( \phi_E^{\mathrm{v}}\left[ n \right] - \phi_E^{\mathrm{v},\left( m \right)}\left[ n \right] \right) \right). \label{FphiV}
\end{align}
\end{subequations}
Then the concave lower bound of ${\hat R}_{U_k}^{\mathrm{sec}}\left[ n \right]$ is obtained as
\begin{align}
	\tilde{R}_{U_k}^{\mathrm{sec,lb,v}}\left[ n \right] &=G_{U_k,1}^{\mathrm{v},\left( m \right)}\left[ n \right] + G_{U_k,2}^{\mathrm{v},\left( m \right)}\left[ n \right] \left( p_{U_k}^{\mathrm{v}}\left[ n \right] - p_{U_k}^{\mathrm{v},\left( m \right)}\left[ n \right] \right) \nonumber \\
	&+ G_{U_k,3}^{\mathrm{v},\left( m \right)}\left[ n \right] \left( \mu_{U_k}^{\mathrm{v}}\left[ n \right] - \mu_{U_k}^{\mathrm{v},\left( m \right)}\left[ n \right] \right) - G_E^{\mathrm{v}}\left[ n \right],
\end{align}
where
\begin{subequations}
\begin{align}
	G_{U_k,1}^{\mathrm{v},\left( m \right)}\left[ n \right] &= \frac{ \log_2 \left( 1 + \gamma_{U_k}^{\mathrm{v},\left( m \right)}\left[ n \right] \right) }{ p_{U_k}^{\mathrm{v},\left( m \right)}\left[ n \right] }, \\
	G_{U_k,2}^{\mathrm{v},\left( m \right)}\left[ n \right] &= -\frac{ \log_2 \left( 1 + \gamma_{U_k}^{\mathrm{v},\left( m \right)}\left[ n \right] \right) }{ \left( p_{U_k}^{\mathrm{v},\left( m \right)}\left[ n \right] \right)^2 }, \\
	G_{U_k,3}^{\mathrm{v},\left( m \right)}\left[ n \right] &= \frac{ - \theta_{U_k}\left[ n \right] S_{U_k}\left[ n \right] / \left( 1 + \gamma_{U_k}^{\mathrm{v},\left( m \right)}\left[ n \right] \right) }{ \ln 2 \cdot p_{U_k}^{\mathrm{v},\left( m \right)}\left[ n \right] \left( I_{U_k}\left[ n \right] + \mu_{U_k}^{\mathrm{v},\left( m \right)}\left[ n \right] \right)^2 }, \\
	\gamma_{U_k}^{\mathrm{v},\left( m \right)}\left[ n \right] &= \frac{ \theta_{U_k}\left[ n \right] S_{U_k}\left[ n \right] }{ I_{U_k}\left[ n \right] + \mu_{U_k}^{\mathrm{v},\left( m \right)}\left[ n \right] }.
	\label{H232}
\end{align}
\end{subequations}

The sensing constraint (\ref{Opt10c}) is transformed into
\begin{align}
	\log_2 \left( 1 + \frac{ \Gamma_{V_j}^{\mathrm{rad}}\left[ n \right] }{ \varphi_{V_j}^{\mathrm{v},\left( m \right)}\left[ n \right] } \right) &- \frac{ \Gamma_{V_j}^{\mathrm{rad}}\left[ n \right] \left( \frac{ \varphi_{V_j}^{\mathrm{v}}\left[ n \right] }{ \varphi_{V_j}^{\mathrm{v},\left( m \right)}\left[ n \right] } - 1 \right) }{ \ln 2 \left( \Gamma_{V_j}^{\mathrm{rad}}\left[ n \right] + \varphi_{V_j}^{\mathrm{v},\left( m \right)}\left[ n \right] \right) } \nonumber \\
	&\geq 2 \theta_{V_j}\left[ n \right] \mu \delta^{-1} R_{\min}^{\mathrm{rad}} p_{V_j}^{\mathrm{v}}\left[ n \right]. \label{C2rz}
\end{align}

The computation constraint (\ref{Opt10d}) is recast as
\begin{align}
	\Gamma^f\left[ n \right] p_B^{\mathrm{v}}\left[ n \right] \leq \tilde{R}_B^{\mathrm{app,v}}\left[ n \right], \label{C3rz}
\end{align}
where $\tilde{R}_B^{\mathrm{app,v}}\left[ n \right] = \log_2 \left( S_B\left[ n \right] + I_B\left[ n \right] + \mu_B^{\mathrm{v}}\left[ n \right] \right) - \log_2 \left( I_B\left[ n \right] + \mu_B^{\mathrm{v},\left( m \right)}\left[ n \right] \right) - \frac{ \mu_B^{\mathrm{v}}\left[ n \right] - \mu_B^{\mathrm{v},\left( m \right)}\left[ n \right] }{ \ln 2 \left( I_B\left[ n \right] + \mu_B^{\mathrm{v},\left( m \right)}\left[ n \right] \right) }$, and $\mu_B^{\mathrm{v}}\left[ n \right] = \beta_0^{-1} \sigma_B^2 \varphi_B^{\mathrm{v}}\left[ n \right]$.

Thus, $\mathcal{P}_{1.4{\rm a}}$ is reformulated as the following convex problem,
\begin{subequations}
\begin{align}
	\mathcal{P}_{1.4{\rm b}}: \quad &\max_{\mathbf{Z}, \mathbf{p}^{\mathrm{v}}, \bm{\varphi}^{\mathrm{v}}} \; \frac{1}{N}\sum\limits_{n = 1}^N \sum\limits_{k = 1}^K {{\tilde{R}}_{U_k}^{\mathrm{sec,lb,v}}\left[ n \right]} \\
	\text{s.t.} \quad & \tilde{R}_{U_k}^{\mathrm{sec,lb,v}}\left[ n \right] \geq \theta_{U_k}\left[ n \right] R_{\min}^{\mathrm{tra}}, \forall k, n, \label{C1rz} \\
	& \mathrm{(\ref{Opt10i}), (\ref{Opt10k})-(\ref{Opt10m}), (\ref{C1a1v})-(\ref{C3a1v}), (\ref{C2rz}), (\ref{C3rz})}. \nonumber
\end{align}
\end{subequations}
which can be efficiently solved using CVX.

Finally, an AO-based algorithm is developed to solve $\mathcal{P}_{1.0}$ by iteratively solving the subproblems discussed above. The overall procedure is summarized in Algorithm 1, where $\varepsilon$ denotes the convergence tolerance.

\begin{algorithm}[tb]
	\color{black}
	\caption{Iterative Algorithm for Problem ($\mathcal{P}_{1.0}$)}\label{algorithm1}
	\KwIn{Initialize feasible points}
	\While
	{$R\left( {{\bm \theta}^{\left( {m} \right)},{\bf F}^{\left( {m} \right)},{{\bf \tilde W}^{\left( {m} \right)}},{\bf Q}_s^{\left( {m} \right)},{{\bf Z}^{\left( {m} \right)}}} \right) - R\left( {{\bm \theta}^{\left( {m - 1} \right)},{\bf F}^{\left( {m - 1} \right)},{{\bf \tilde W}^{\left( {m - 1} \right)}},{\bf Q}_s^{\left( {m - 1} \right)},{{\bf Z}^{\left( {m - 1} \right)}}} \right) > \varepsilon $}
	{1. Obtain the solution ${\bm \theta}^{\left( {m + 1} \right)}$ by solving ($\mathcal{P}_{1.1}$) for given $\left\{ {\bf F}^{\left( {m} \right)},{{\bf \tilde W}^{\left( {m} \right)}},{\bf Q}_s^{\left( {m} \right)},{{\bf Z}^{\left( {m} \right)}} \right\}$;\\
		2. Obtain the solution ${\bf F}^{\left( {m + 1} \right)}, {{\bf \tilde W}^{\left( {m + 1} \right)}}$ by solving ($\mathcal{P}_{1.2}$) for given $\left\{ {{\bm \theta}^{\left( {m + 1} \right)},{\bf Q}_s^{\left( m \right)},{\bf Z}^{\left( {m} \right)}} \right\}$;\\
		3. Obtain the solution ${\bf Q}_s^{\left( {m + 1} \right)}$ by solving ($\mathcal{P}_{1.3}$) for given $\left\{ {{\bm \theta}^{\left( {m + 1} \right)},{\bf F}^{\left( {m + 1} \right)},{\bf \tilde W}^{\left( m+1 \right)},{\bf Z}^{\left( {m} \right)} } \right\}$;\\
		4. Obtain the solution ${\bf Z}^{\left( {m + 1} \right)}$ by solving ($\mathcal{P}_{1.4}$) for given $\left\{ {{\bm \theta}^{\left( {m + 1} \right)},{\bf F}^{\left( {m + 1} \right)},{{\bf \tilde W}^{\left( {m + 1} \right)}},{\bf Q}_s^{\left( {m + 1} \right)}} \right\}$;\\
		5. $m = m + 1$;\\
		6. Compute the objective value $R\left( {{\bm \theta}^{\left( {m} \right)},{\bf F}^{\left( {m} \right)},{{\bf \tilde W}^{\left( {m} \right)}},{\bf Q}_s^{\left( {m} \right)},{{\bf Z}^{\left( {m} \right)}}} \right)$.\\
	}
	\KwOut{${\bar{\mathcal{R}}}^{\mathrm{sec}}$}
\end{algorithm}

\section{Sensing-centric Design}
\label{sec:SensingCen}

In this section, we consider sensing-centric scenarios, such as environmental monitoring and agricultural Internet of Things (IoT) applications, where the average radar estimation rate (ARER) is maximized through joint optimization of scheduling, computational frequency, beamforming, and UAV trajectory. The following optimization problem is formulated 
\begin{subequations}
	\begin{align}
		\mathcal{P}_{2.0}: \quad &\max_{\bm{\theta}, \mathbf{F}, \tilde{\mathbf{W}}, \mathbf{Q}_s, \mathbf{Z}} \; {\bar{\mathcal{R}}^{\mathrm{sen}}} \\
		\text{s.t.} \quad & \mathrm{(\ref{scheduling1a}) - (\ref{scheduling2b})}, \mathrm{(\ref{Opt10c})-(\ref{Opt10m}), (\ref{P1.1b})}. \nonumber
	\end{align}
\end{subequations}
where 
${\bar{\mathcal{R}}^{\mathrm{sen}}} = \frac{1}{N}\sum\limits_{n = 1}^N {\sum\limits_{j = 1}^J {R_{{V_j}}^{{\rm{rad}}}} } \left[ n \right]$ denotes the ARER.

\subsection{Communication and Sensing Scheduling Optimization}

In this subsection, we optimize the scheduling variables $\bm{\theta}$ with $\{\mathbf{F}, \tilde{\mathbf{W}}, \mathbf{Q}_s, \mathbf{Z}\}$ fixed. Following a similar procedure as in $\mathcal{P}_{1.1}$, the optimization problem is reformulated as
\begin{subequations}
\begin{align}
	\mathcal{P}_{2.1}: \quad &\max_{\bm{\theta}} \; {\bar{\mathcal{R}}^{\mathrm{sen}}}\\
	\text{s.t.} \quad & \mathrm{\mathrm{(\ref{scheduling1a}), (\ref{scheduling2a})}, (\ref{Opt10d}), \mathrm{(\ref{P1.1b})}-\mathrm{(\ref{scheduling2b2}), (\ref{P1.1C2})}}. \nonumber
\end{align}
\end{subequations}
$\mathcal{P}_{2.1}$ is convex and can be efficiently solved using CVX.

\subsection{Computation and Power Allocation Optimization}

In this subsection, we optimize the computational frequency $\mathbf{F}$ and the beamforming matrices $\tilde{\mathbf{W}}$ with $\{\bm{\theta}, \mathbf{Q}_s, \mathbf{Z}\}$ fixed. Similar to $\mathcal{P}_{1.2}$, by dropping the rank-one constraint, $\mathcal{P}_{2.0}$ is reformulated as
\begin{subequations}
\begin{align}
	\mathcal{P}_{2.2}: \quad &\max_{\mathbf{F}, \tilde{\mathbf{W}}} \; {\bar{\mathcal{R}}^{\mathrm{sen}}}\\
	\text{s.t.} \quad & \mathrm{(\ref{Opt10e}), (\ref{Opt10f}), (\ref{Opt10h}), (\ref{Opt10i}), (\ref{P1.1C2}), (\ref{P1.2C1}), (\ref{P1.2C3})}. \nonumber
\end{align}
\end{subequations}
$\mathcal{P}_{2.2}$ is convex and can be solved using CVX. Subsequently, standard techniques such as Gaussian randomization can be applied to recover a rank-one solution.

\subsection{Horizontal Trajectory Optimization of $S$}
In this subsection, we optimize the horizontal trajectory $\mathbf{Q}_s$ with $\{\bm{\theta}, \mathbf{F}, \tilde{\mathbf{W}}, \mathbf{Z}\}$ fixed. Accordingly, $\mathcal{P}_{2.0}$ is reformulated as
\begin{subequations}
\begin{align}
	\mathcal{P}_{2.3{\rm a}}: \quad &\max_{\mathbf{Q}_s} \; {\bar{\mathcal{R}}^{\mathrm{sen}}} \label{P2.3a}\\
	\text{s.t.} \quad & \mathrm{(\ref{Opt10c})-(\ref{Opt10d}), (\ref{Opt10i}), (\ref{Opt10j}), (\ref{Opt10m}), (\ref{P1.1b})}. \nonumber
\end{align}
\end{subequations}
In $\mathcal{P}_{2.3{\rm a}}$, the objective function (\ref{P2.3a}) along with constraints (\ref{Opt10c}) - (\ref{Opt10d}), (\ref{Opt10i}), and (\ref{P1.1b}) are all non-convex with respect to $\mathbf{Q}_s$, rendering $\mathcal{P}_{2.3{\rm a}}$ a non-convex optimization problem.

Since $R_{V_j}^{\mathrm{rad,lb}}\left[ n \right]$ in (\ref{P2.3a}) and (\ref{Opt10c}) are non-concave with respect to $\mathbf{Q}_s$, the following relaxed convex function is obtained as
\begin{align}
	R_{V_j}^{\mathrm{rad,lb}}\left[ n \right] = \frac{ \delta \theta_{V_j}\left[ n \right] }{ 2\mu p_{V_j}^{\mathrm{h}}\left[ n \right] } \log_2 \left( 1 + \frac{ \Gamma_{V_j}^{\mathrm{rad}}\left[ n \right] }{ \varphi_{V_j}^{\mathrm{h},2}\left[ n \right] } \right).
	\label{H232}
\end{align}
where 
$\varphi_{V_j}^{\mathrm{h},2}\left[ n \right]$ is a slack variable, defined as (\ref{C2a1h}).
Applying a first-order Taylor expansion to $R_{V_j}^{\mathrm{rad,lb}}\left[ n \right]$ yields the following concave lower bound
\begin{align}
	\tilde{R}_{V_j}^{\mathrm{rad,lb}}\left[ n \right] & =  G_{V_j,1}^{\mathrm{h},\left( m \right)}\left[ n \right] + G_{V_j,2}^{\mathrm{h},\left( m \right)}\left[ n \right] \left( p_{V_j}^{\mathrm{h}}\left[ n \right] - p_{V_j}^{\mathrm{h},\left( m \right)}\left[ n \right] \right) \nonumber \\
	&+ G_{V_j,3}^{\mathrm{h},\left( m \right)}\left[ n \right] \left( \varphi_{V_j}^{\mathrm{h},2}\left[ n \right] - \varphi_{V_j}^{\mathrm{h},2,\left( m \right)}\left[ n \right] \right),
\end{align}
where 
\begin{subequations}
	\begin{align}
		G_{V_j,1}^{\mathrm{h},\left( m \right)}\left[ n \right] &= \frac{ \delta \theta_{V_j}\left[ n \right] }{ 2\mu p_{V_j}^{\mathrm{h},\left( m \right)}\left[ n \right] } \log_2 \left( 1 + \gamma_{V_j}^{\mathrm{h},\left( m \right)}\left[ n \right] \right), \label{Ghmv1} \\
		G_{V_j,2}^{\mathrm{h},\left( m \right)}\left[ n \right] &= - \frac{ \delta \theta_{V_j}\left[ n \right] }{ 2\mu \left( p_{V_j}^{\mathrm{h},\left( m \right)}\left[ n \right] \right)^2 } \log_2 \left( 1 + \gamma_{V_j}^{\mathrm{h},\left( m \right)}\left[ n \right] \right), \label{Ghmv2}\\
		G_{V_j,3}^{\mathrm{h},\left( m \right)}\left[ n \right] &= - \frac{ \delta \theta_{V_j}\left[ n \right] \gamma_{V_j}^{\mathrm{h},\left( m \right)}\left[ n \right] }{ 2 \ln 2 \cdot \mu p_{V_j}^{\mathrm{h},\left( m \right)}\left[ n \right] \left( \Gamma_{V_j}^{\mathrm{rad}}\left[ n \right] + \varphi_{V_j}^{\mathrm{h},2,\left( m \right)}\left[ n \right] \right) }, \label{Ghmv3}\\
		\gamma_{V_j}^{\mathrm{h},\left( m \right)}\left[ n \right] &= \frac{ \Gamma_{V_j}^{\mathrm{rad}}\left[ n \right] }{ \varphi_{V_j}^{\mathrm{h},2,\left( m \right)}\left[ n \right] }.\label{yhmv} 
	\end{align}
\end{subequations}

Following the same procedure as in $\mathcal{P}_{1.3{\rm a}}$, $\mathcal{P}_{2.3{\rm a}}$ is reformulated as
\begin{subequations}
\begin{align}
	\mathcal{P}_{2.3{\rm b}}: \quad &\max_{\mathbf{Q}_s, \mathbf{p}^{\mathrm{h}}, \bm{\varphi}^{\mathrm{h}}, \mathbf{v}} \; \frac{1}{N} \sum_{n=1}^{N} \sum_{j=1}^{J} \tilde{R}_{V_j}^{\mathrm{rad,lb}}\left[ n \right] \\
	\text{s.t.} \quad & \tilde{R}_{V_j}^{\mathrm{rad,lb}}\left[ n \right] \geq \theta_{V_j}\left[ n \right] R_{\min}^{\mathrm{rad}}, \forall j, n \label{C2rq2} \\
	& \mathrm{(\ref{Opt10j}), (\ref{Opt10m}), (\ref{C1a1h}), (\ref{C1a4h}), (\ref{C1a5h})}, \nonumber \\
	& \mathrm{(\ref{C1a2h})-(\ref{C1a6h}), (\ref{C1a7h}), (\ref{C1a8h})}, \nonumber \\
	& \mathrm{(\ref{C2a1h}) - (\ref{v1}), (\ref{v2t}), (\ref{C1rq})}. \nonumber
\end{align}
\end{subequations}
$\mathcal{P}_{2.3{\rm b}}$ is convex and can be efficiently solved using CVX.

\subsection{Vertical Trajectory Optimization of $S$}

In this subsection, we optimize the vertical trajectory $\mathbf{Z}$ with $\{\bm{\theta}, \mathbf{F}, \tilde{\mathbf{W}}, \mathbf{Q}_s\}$ fixed. The optimization problem is formulated as
\begin{subequations}
	\begin{align}
		\mathcal{P}_{2.4{\rm a}}: \quad &\max_{\mathbf{Z}, \mathbf{p}^{\mathrm{v}}, \bm{\varphi}^{\mathrm{v}}} \; \frac{1}{N} \sum_{n=1}^{N} \sum_{j=1}^{J} \tilde{R}_{V_j}^{\mathrm{rad,lb}}\left[ n \right] \label{P2.4a} \\
		\text{s.t.} \quad & \mathrm{(\ref{Opt10i}), (\ref{Opt10k})-(\ref{Opt10m}), (\ref{C1a1v})-(\ref{C3a1v})}, \nonumber \\
		& \mathrm{(\ref{C3rz}), (\ref{C1rz}), (\ref{C2rq2})}. \nonumber
	\end{align}
\end{subequations}
where $G_{V_j,1}^{\mathrm{v},\left( m \right)}\left[ n \right]$, $G_{V_j,2}^{\mathrm{v},\left( m \right)}\left[ n \right]$, $G_{V_j,3}^{\mathrm{v},\left( m \right)}\left[ n \right]$, and $\gamma_{V_j}^{\mathrm{v},\left( m \right)}\left[ n \right]$ are defined analogously to their horizontal counterparts (\ref{Ghmv1}), (\ref{Ghmv2}), (\ref{Ghmv3}) and (\ref{yhmv}), with the superscript $\mathrm{h}$ replaced by $\mathrm{v}$.
$\mathcal{P}_{2.4{\rm a}}$ is convex and can be solved using CVX.

Finally, similar to $\mathcal{P}_{1.0}$, an AO-based algorithm can be developed to solve $\mathcal{P}_{2.0}$ by iteratively solving the above subproblems. Due to space limitations, the detailed procedure is omitted here.

\section{Computing-Centric Design}
\label{sec:ComputingCen}

This section considers a computation-centric scenario, such as model training and massive data processing, where the average computing energy efficiency is maximized through joint optimization of scheduling, computational frequency, beamforming, and UAV trajectory. The optimization problem is formulated as
\begin{subequations}
\begin{align}
	\mathcal{P}_{3.0}: \quad &\max_{\bm{\theta}, \mathbf{F}, \tilde{\mathbf{W}}, \mathbf{Q}_s, \mathbf{Z}} \; {\bar \Phi}  \\
	\text{s.t.} \quad & \mathrm{(\ref{scheduling1a})-(\ref{scheduling2b}), (\ref{Opt10c})-(\ref{Opt10m}), (\ref{P1.1b})}. \nonumber
\end{align}
\end{subequations}
where 
${\bar \Phi}  = \frac{1}{N}\sum\limits_{n = 1}^N {\frac{{{D_J}\left[ n \right]}}{{{E_s}\left[ n \right]}}}$ denotes the average computing energy efficiency and 
$E_s\left[ n \right] = \alpha_1 E^{\mathrm{cop}}\left[ n \right] + \alpha_2 E^{\mathrm{tra}}\left[ n \right] + \alpha_3 E^{\mathrm{fly}}\left[ n \right]$ represents the weighted total energy consumption. Due to the dominance of UAV propulsion energy, weighting coefficients $\alpha_1$, $\alpha_2$, and $\alpha_3$ are introduced to balance the three energy components.

\subsection{Communication and Sensing Scheduling Optimization}

In this subsection, we optimize the scheduling variables $\bm{\theta}$ with $\{\mathbf{F}, \tilde{\mathbf{W}}, \mathbf{Q}_s, \mathbf{Z}\}$ fixed. Following a similar procedure as in $\mathcal{P}_{1.1}$, the optimization problem is reformulated as
\begin{subequations}
\begin{align}
	\mathcal{P}_{3.1}: \quad &\max_{\bm{\theta}} \; {\bar \Phi} \\
	\text{s.t.} \quad & \mathrm{\mathrm{(\ref{scheduling1a}), (\ref{scheduling2a})}, (\ref{Opt10d}), \mathrm{(\ref{P1.1b})}-\mathrm{(\ref{scheduling2b2}), (\ref{P1.1C2})}}. \nonumber
\end{align}
\end{subequations}
$\mathcal{P}_{3.1}$ is convex and can be efficiently solved using CVX.

\subsection{Computation and Power Allocation Optimization}

In this subsection, we optimize the computational frequency $\mathbf{F}$ and the beamforming matrices $\tilde{\mathbf{W}}$ with $\{\bm{\theta}, \mathbf{Q}_s, \mathbf{Z}\}$ fixed. Similar to $\mathcal{P}_{1.2}$, by dropping the rank-one constraint, $\mathcal{P}_{3.2}$ is reformulated as
\begin{subequations}
	\begin{align}
		\mathcal{P}_{3.2{\rm a}}: \quad &\max_{\mathbf{F}, \tilde{\mathbf{W}}} \; {\bar \Phi} \label{P3.2a}\\
		\text{s.t.} \quad & \mathrm{(\ref{Opt10c})-(\ref{Opt10f}), (\ref{Opt10h}), (\ref{Opt10i}), (\ref{P1.1b})}. \nonumber
	\end{align}
\end{subequations}
In $\mathcal{P}_{3.2{\rm a}}$, the objective function (\ref{P3.2a}) is non-convex with respect to $\tilde{\mathbf{W}}$, rendering $\mathcal{P}_{3.2{\rm a}}$ a non-convex optimization problem.

Since $E_s\left[ n \right]$ is a complex function of the optimization variables, the fractional term $\frac{D_J\left[ n \right]}{E_s\left[ n \right]}$ is non-concave. To address this issue, we introduce two slack variables $\chi\left[ n \right]$ and $\zeta\left[ n \right]$ that satify
\begin{subequations}
	\begin{align}	
		\chi\left[ n \right] &\le \frac{D_J\left[ n \right]}{E_s\left[ n \right]}, \label{C15m}\\
		\zeta\left[ n \right] &\ge E_s\left[ n \right]. \label{C16} 
	\end{align}
\end{subequations}
Here, $\zeta\left[ n \right]$ serves as an upper bound of $E_s\left[ n \right]$ and $\chi\left[ n \right]$ serves as a lower bound of the objective function $\frac{D_J\left[ n \right]}{E_s\left[ n \right]}$. Then we have 
\begin{align}
	\chi\left[ n \right] \zeta\left[ n \right] \le D_J\left[ n \right]. \label{C15m2}
\end{align}
It should be noted that 
the non-convex product $\chi\left[ n \right] \zeta\left[ n \right]$ make (\ref{C15m2}) no-convex. 
Following \cite{WangW2022JSAC}, we employ a convex upper bound approximation deal with (\ref{C15m}) and obtain
\begin{align}
	\left( \chi\left[ n \right] \zeta\left[ n \right] \right)^{\mathrm{ub}} \leq D_J\left[ n \right], \label{C15}
\end{align}
where 
$\left( \chi\left[ n \right] \zeta\left[ n \right] \right)^{\mathrm{ub}} = \frac{t^{\left( m \right)}\left[ n \right]}{2} \chi^2\left[ n \right] + \frac{ \zeta^2\left[ n \right] }{ 2 t^{\left( m \right)}\left[ n \right] }$ with $t^{\left( m \right)}\left[ n \right] = \zeta^{(m-1)}\left[ n \right] / \chi^{(m-1)}\left[ n \right]$. Consequently, $\mathcal{P}_{3.2{\rm a}}$ is reformulated as
\begin{subequations}
	\begin{align}
		\mathcal{P}_{3.2{\rm b}}: \quad &\max_{\mathbf{F}, \tilde{\mathbf{W}}, \bm{\Lambda}} \; \Psi  \\
		\text{s.t.} \quad & \mathrm{(\ref{Opt10e}), (\ref{Opt10f}), (\ref{Opt10h}), (\ref{Opt10i}), (\ref{P1.1C2})}, \nonumber\\
		& \mathrm{(\ref{P1.2C1}), (\ref{P1.2C3}), (\ref{C16}), (\ref{C15})}. \nonumber
	\end{align}
\end{subequations}
where 
$\Psi  = \frac{1}{N} \sum_{n=1}^{N} \chi\left[ n \right]$ 
and 
$\bm{\Lambda} = \{ \chi\left[ n \right], \zeta\left[ n \right], \forall n \}$. 
$\mathcal{P}_{3.2{\rm b}}$ is convex and can be solved using CVX. Subsequently, Gaussian randomization can be applied to recover a rank-one solution.

\subsection{Horizontal Trajectory Optimization of $S$}

In this subsection, we optimize the horizontal trajectory $\mathbf{Q}_s$ with $\{\bm{\theta}, \mathbf{F}, \tilde{\mathbf{W}}, \mathbf{Z}\}$ fixed. Following a similar procedure as in $\mathcal{P}_{1.3}$ and $\mathcal{P}_{2.3}$, $\mathcal{P}_{3.0}$ is reformulated as
\begin{subequations}
	\begin{align}
		\mathcal{P}_{3.3}: \quad &\max_{\mathbf{Q}_s, \mathbf{p}^{\mathrm{h}}, \bm{\varphi}^{\mathrm{h}}, \mathbf{v}, \bm{\Lambda}} \; \Psi \\
		\text{s.t.} \quad & \mathrm{(\ref{Opt10j}), (\ref{Opt10m}), (\ref{C1a1h}), (\ref{C1a4h}), (\ref{C1a5h})}, \nonumber \\
		& \mathrm{(\ref{C1a2h})-(\ref{C1a6h}), (\ref{C1a7h}), (\ref{C1a8h})}, \nonumber \\
		& \mathrm{(\ref{C2a1h}) - (\ref{v1}), (\ref{v2t}), (\ref{C1rq}), (\ref{C2rq2}), (\ref{C16}), (\ref{C15})}. \nonumber
	\end{align}
\end{subequations}
$\mathcal{P}_{3.3}$ is convex and can be efficiently solved using CVX.

\subsection{Vertical Trajectory Optimization of $S$}
In this subsection, we optimize the vertical trajectory $\mathbf{Z}$ with $\{\bm{\theta}, \mathbf{F}, \tilde{\mathbf{W}}, \mathbf{Q}_s\}$ fixed. Following a similar procedure as in $\mathcal{P}_{1.4}$ and $\mathcal{P}_{3.2}$, $\mathcal{P}_{3.0}$ is reformulated as
\begin{subequations}
	\begin{align}
		\mathcal{P}_{3.4}: \quad &\max_{\mathbf{Z}, \mathbf{p}^{\mathrm{v}}, \bm{\varphi}^{\mathrm{v}}, \bm{\Lambda}} \; \Psi \\
		\text{s.t.} \quad & \mathrm{(\ref{Opt10i}), (\ref{Opt10k})-(\ref{Opt10m}), (\ref{C1a1v})-(\ref{C3a1v})}, \nonumber \\
		& \mathrm{(\ref{C3rz}), (\ref{C1rz}), (\ref{C2rq2}), (\ref{C16}), (\ref{C15})}. \nonumber
	\end{align}
\end{subequations}
$\mathcal{P}_{3.4}$ is convex and can be solved using CVX.

Finally, similar to $\mathcal{P}_{1.0}$, an AO-based algorithm can be developed to solve $\mathcal{P}_{3.0}$ by iteratively solving the above subproblems. Due to space limitations, the detailed procedure is omitted here.

\begin{algorithm}[tb]
	\color{black}
	\caption{Iterative Algorithm for Problem ($\mathcal{P}_{4.0}$)}\label{algorithm2}
	\KwIn{${\tilde{\mathcal{R}}^{\mathrm{sec}}}$, ${\tilde{\mathcal{R}}^{\mathrm{sen}}}$, $\tilde{{\Phi}}$, and initial feasible points}
	\While
	{$\Xi\left( \bm{\theta}^{\left( m \right)}, \mathbf{F}^{\left( m \right)}, \tilde{\mathbf{W}}^{\left( m \right)}, \mathbf{Q}_s^{\left( m \right)}, \mathbf{Z}^{\left( m \right)} \right) - \Xi\left( \bm{\theta}^{(m-1)}, \mathbf{F}^{(m-1)}, \tilde{\mathbf{W}}^{(m-1)}, \mathbf{Q}_s^{(m-1)}, \mathbf{Z}^{(m-1)} \right) > \varepsilon $}
	{1. Obtain $\bm{\theta}^{(m+1)}$ by solving $\mathcal{P}_{4.1}$ with $\{\mathbf{F}^{\left( m \right)}, \tilde{\mathbf{W}}^{\left( m \right)}, \mathbf{Q}_s^{\left( m \right)}, \mathbf{Z}^{\left( m \right)}\}$ fixed;\\
		2. Obtain $\mathbf{F}^{(m+1)}, \tilde{\mathbf{W}}^{(m+1)}$ by solving $\mathcal{P}_{4.2}$ with $\{\bm{\theta}^{(m+1)}, \mathbf{Q}_s^{\left( m \right)}, \mathbf{Z}^{\left( m \right)}\}$ fixed;\\
		3. Obtain $\mathbf{Q}_s^{(m+1)}$ by solving $\mathcal{P}_{4.3}$ with $\{\bm{\theta}^{(m+1)}, \mathbf{F}^{(m+1)}, \tilde{\mathbf{W}}^{(m+1)}, \mathbf{Z}^{\left( m \right)}\}$ fixed;\\
		4. Obtain $\mathbf{Z}^{(m+1)}$ by solving $\mathcal{P}_{4.4}$ with $\{\bm{\theta}^{(m+1)}, \mathbf{F}^{(m+1)}, \tilde{\mathbf{W}}^{(m+1)}, \mathbf{Q}_s^{(m+1)}\}$ fixed;\\
		5. $m = m + 1$;\\
		6. Compute the objective value $\Xi\left( \bm{\theta}^{\left( m \right)}, \mathbf{F}^{\left( m \right)}, \tilde{\mathbf{W}}^{\left( m \right)}, \mathbf{Q}_s^{\left( m \right)}, \mathbf{Z}^{\left( m \right)} \right)$.\\
	}
	\KwOut{$\Xi\left( \bm{\theta}^{\left( m \right)}, \mathbf{F}^{\left( m \right)}, \tilde{\mathbf{W}}^{\left( m \right)}, \mathbf{Q}_s^{\left( m \right)}, \mathbf{Z}^{\left( m \right)} \right)$ with $\bm{\theta}^* = \bm{\theta}^{\left( m \right)}$, $\mathbf{F}^* = \mathbf{F}^{\left( m \right)}$, $\tilde{\mathbf{W}}^* = \tilde{\mathbf{W}}^{\left( m \right)}$, $\mathbf{Q}_s^* = \mathbf{Q}_s^{\left( m \right)}$, $\mathbf{Z}^* = \mathbf{Z}^{\left( m \right)}$.}
\end{algorithm}

\begin{figure*}[t]
	\centering
	\subfigure[]{
		\label{fig02a}
		\includegraphics[width = 0.3  \textwidth]{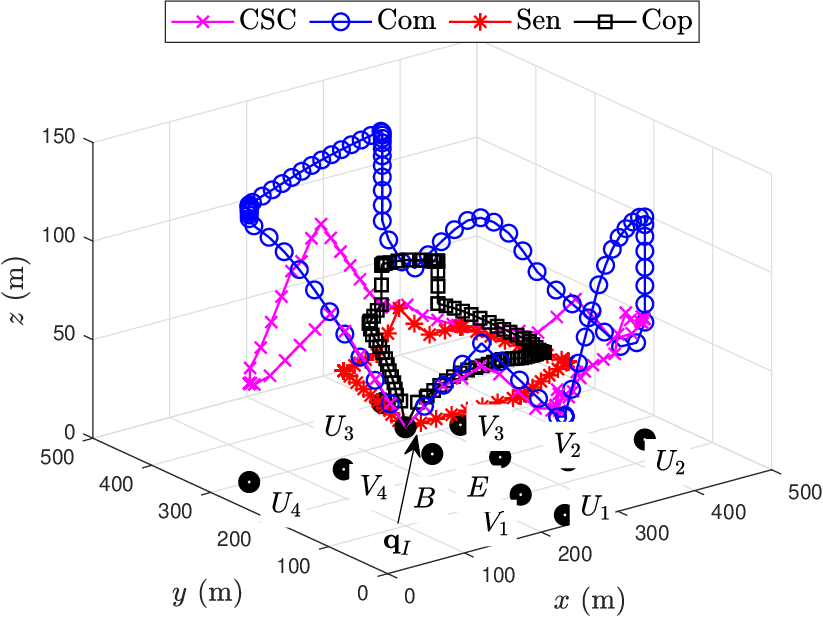}}
	\subfigure[]{
		\label{fig02b}
		\includegraphics[width = 0.3  \textwidth]{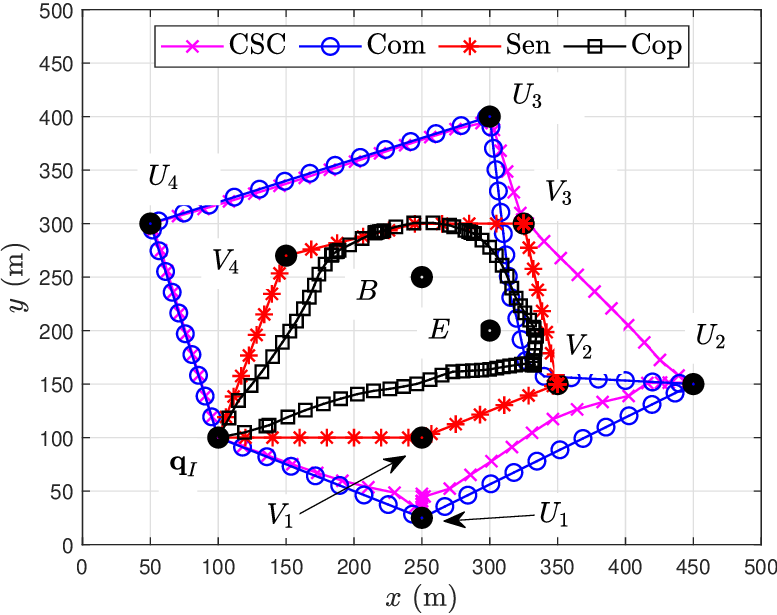}}
	\subfigure[]{
		\label{fig02c}
		\includegraphics[width = 0.3  \textwidth]{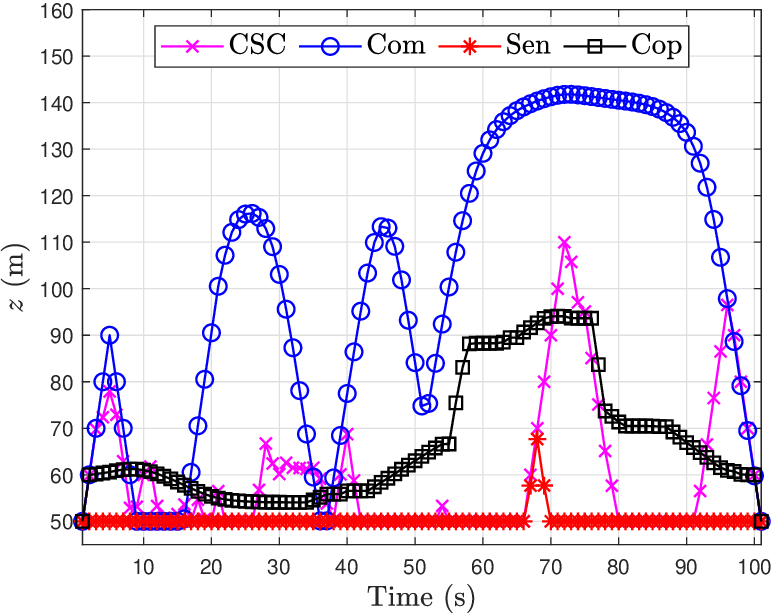}}
	\subfigure[]{
		\label{fig02d}
		\includegraphics[width = 0.3  \textwidth]{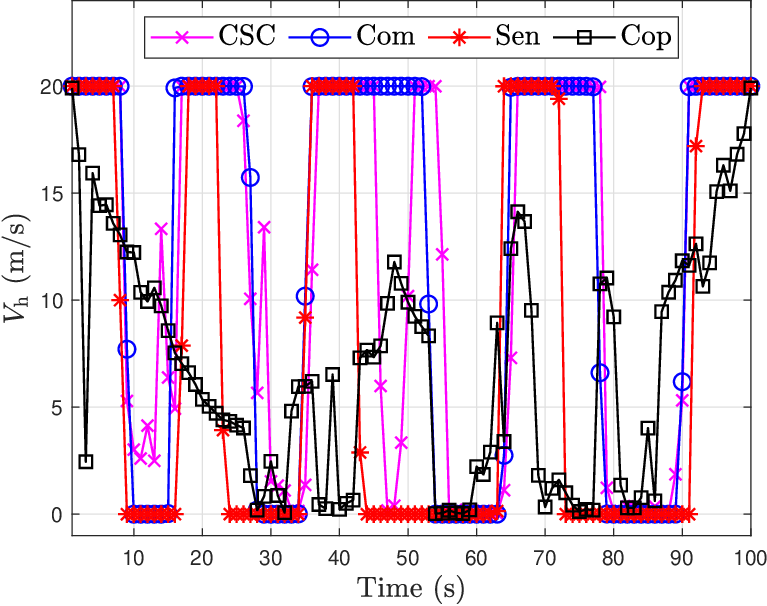}}
	\subfigure[]{
		\label{fig02e}
		\includegraphics[width = 0.3  \textwidth]{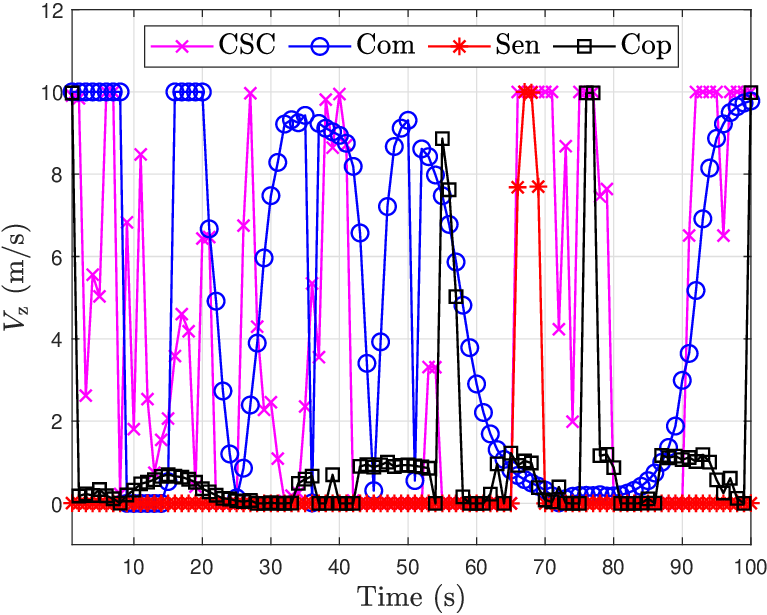}}	
	\subfigure[]{
		\label{fig02f}
		\includegraphics[width = 0.3  \textwidth]{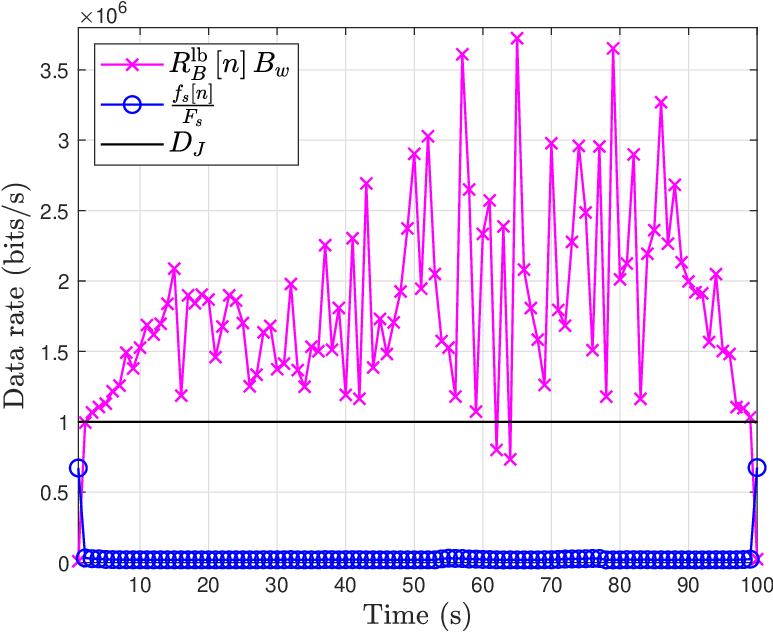}}	
	\caption{Optimized parameters and performance of the four schemes. (a) 3D trajectory of $S$. (b) The horizontal trajectory of $S$. (c) The vertical altitude of $S$. (d) The horizontal speed of $S$. (e) The vertical speed of $S$. (f) Local computing and data offloading capabilities in Cop.}
	\label{fig02}
\end{figure*}
\begin{figure*}[t]
	\centering
	\subfigure[]{
		\label{fig03a}
		\includegraphics[width = 0.3  \textwidth]{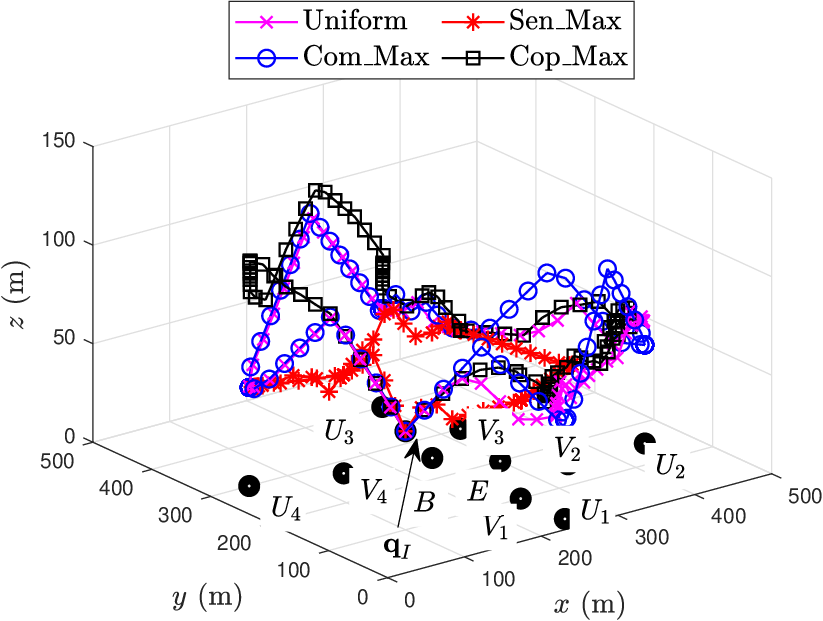}}
	\subfigure[]{
		\label{fig03b}
		\includegraphics[width = 0.3  \textwidth]{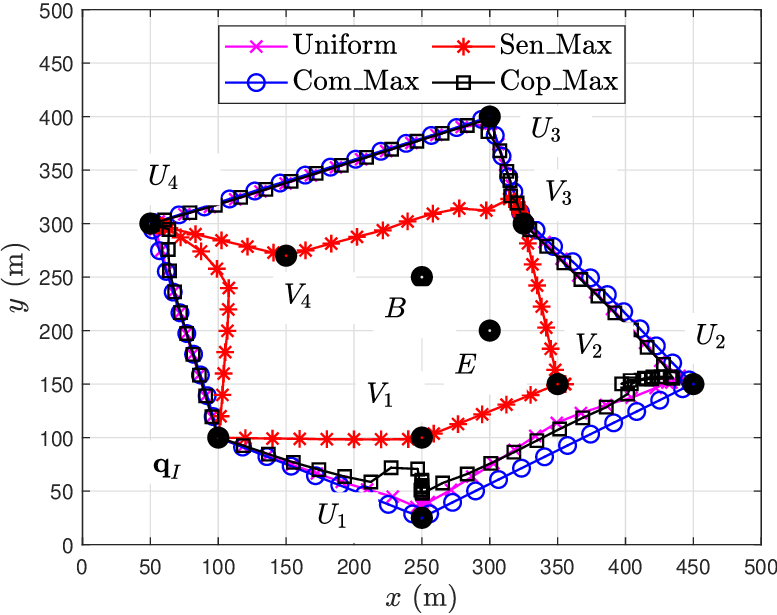}}
	\subfigure[]{
		\label{fig03c}
		\includegraphics[width = 0.3  \textwidth]{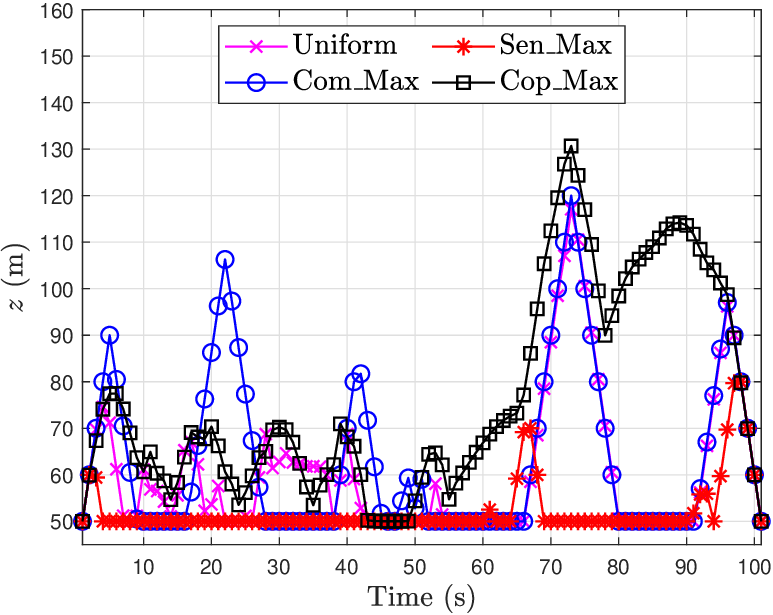}}
	\subfigure[]{
		\label{fig03d}
		\includegraphics[width = 0.3  \textwidth]{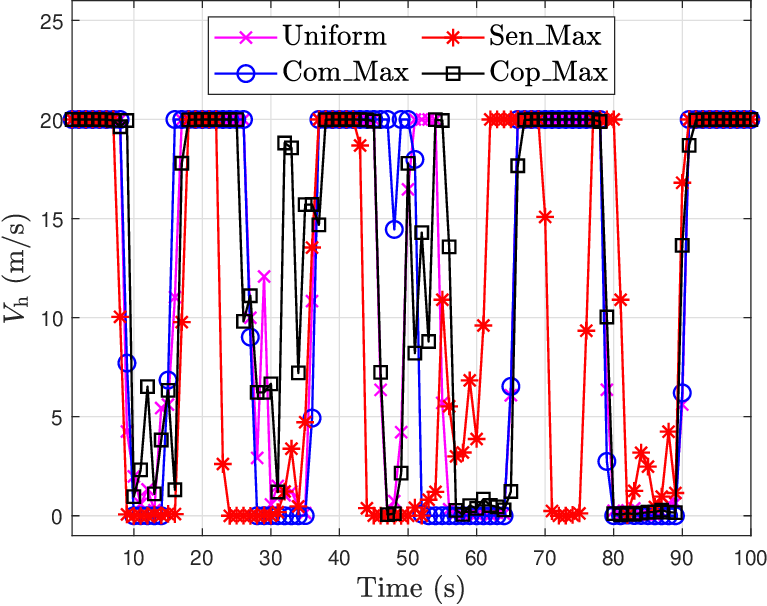}}
	\subfigure[]{
		\label{fig03e}
		\includegraphics[width = 0.3  \textwidth]{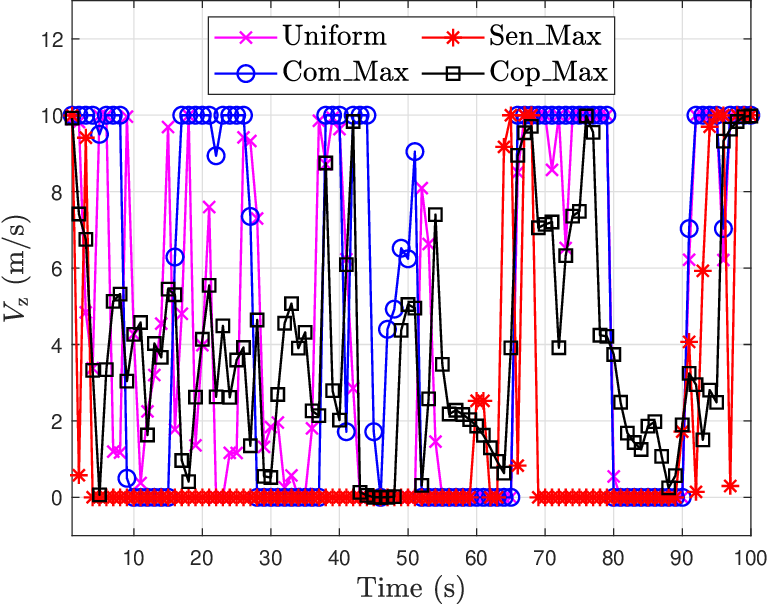}}	
	\subfigure[]{
		\label{fig03f}
		\includegraphics[width = 0.3  \textwidth]{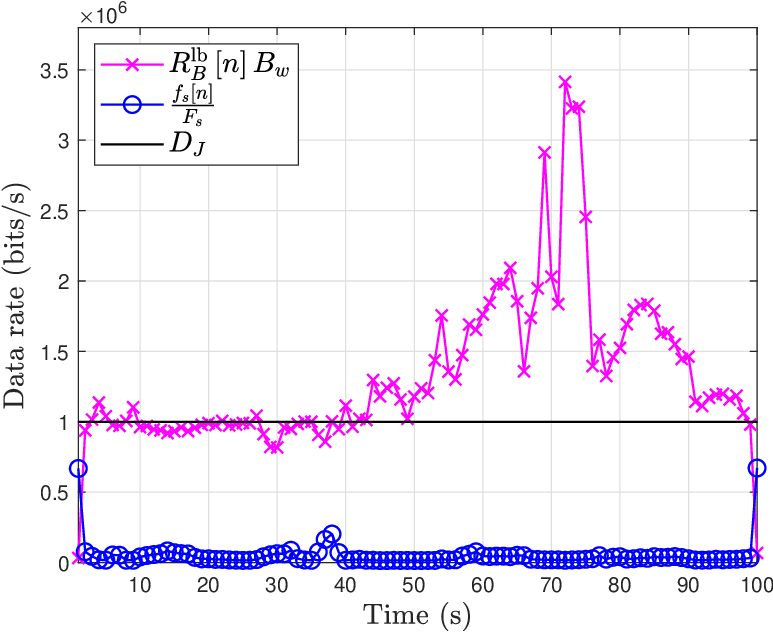}}	
	\caption{Optimized parameters and performance of the CSC scheme with varying weighting coefficients. (a) 3D trajectory of $S$. (b) The horizontal trajectory of $S$. (c) The vertical altitude of $S$. (d) The horizontal speed of $S$. (e) The vertical speed of $S$. (f) Local computing and data offloading capabilities in Cop\_Max. 	}
	\label{fig03}
\end{figure*}
\begin{figure*}[t]
	\centering
	\subfigure[]{
		\label{fig04a}
		\includegraphics[width = 0.3\textwidth]{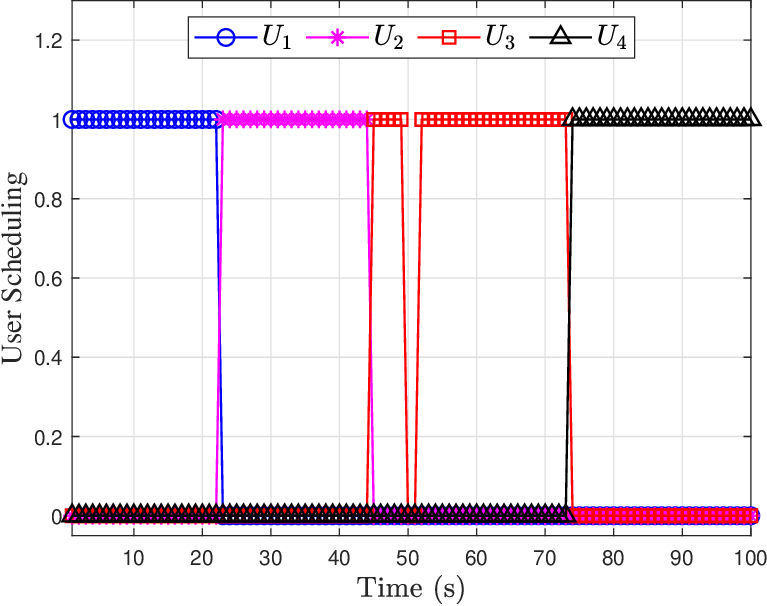}}
	\subfigure[]{
		\label{fig04b}
		\includegraphics[width = 0.3\textwidth]{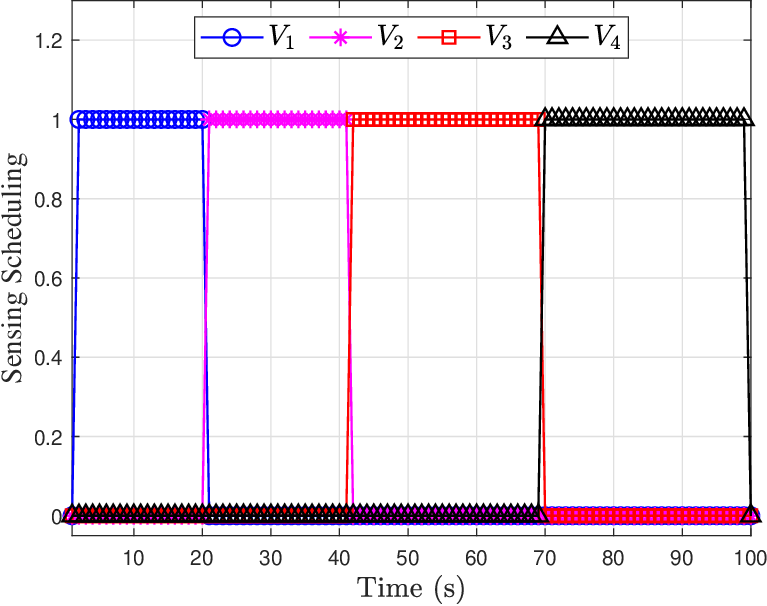}}
	\subfigure[]{
		\label{fig04c}
		\includegraphics[width = 0.3\textwidth]{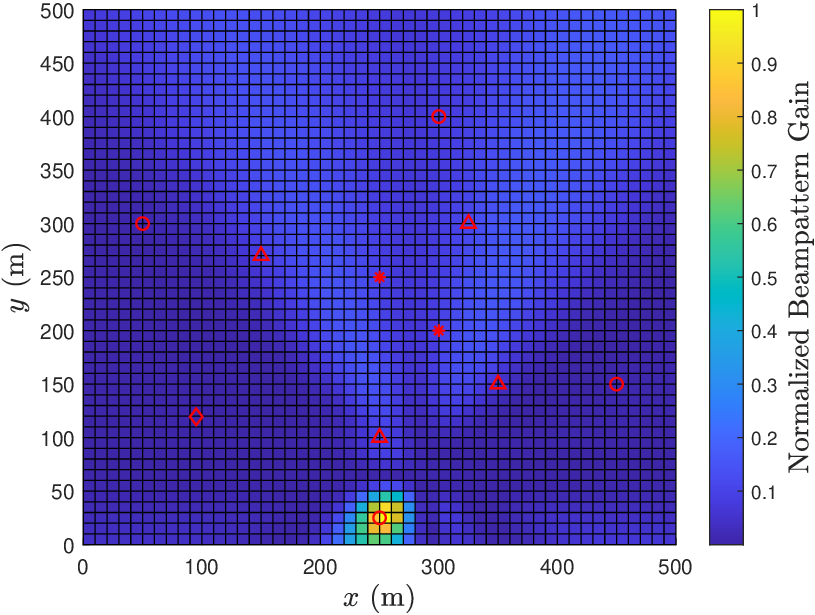}}
	\subfigure[]{
		\label{fig04d}
		\includegraphics[width = 0.3\textwidth]{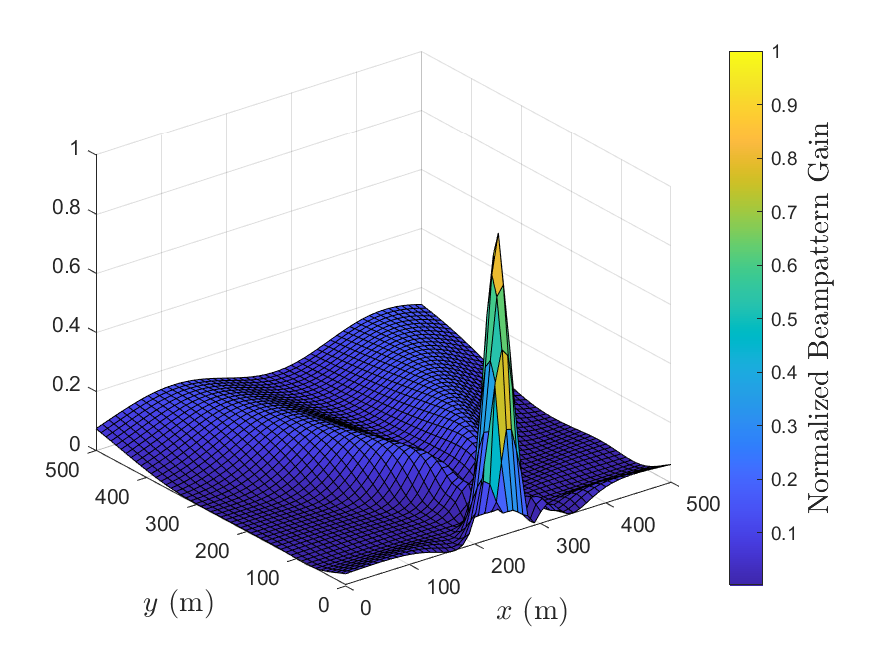}}
	\subfigure[]{
		\label{fig04e}
		\includegraphics[width = 0.3\textwidth]{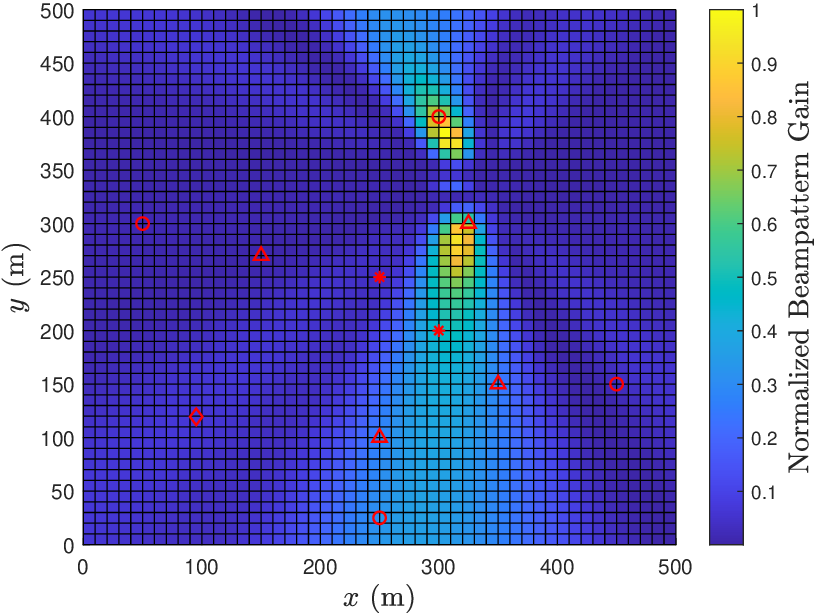}}
	\subfigure[]{
		\label{fig04f}
		\includegraphics[width = 0.3\textwidth]{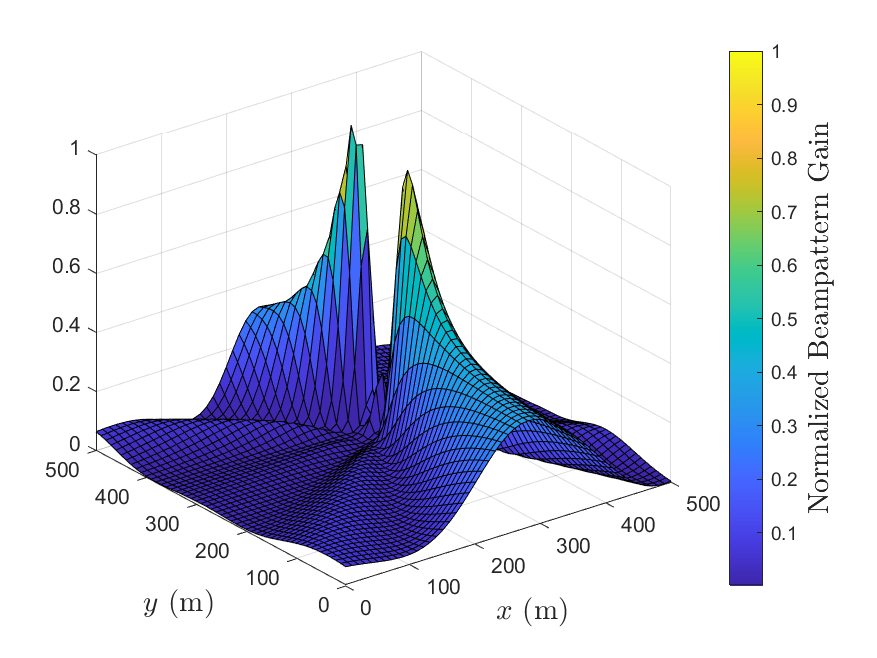}}	
	\caption{The scheduling results and beampattern gains for Uniform.  (a) User scheduling. (b) Target scheduling. (c) 2D normalized beampattern at the $18$-th slot. (d) 3D normalized beampattern at $n = 18$. (e) 2D normalized beampattern at $n = 52$. (f) 3D normalized beampattern at $n = 52$.}
	\label{fig04}
\end{figure*}

\section{Weighted Optimization for Communication, Sensing, and Computing}
\label{sec:WeightedOptimization}

In this section, we explore the trade-off among secure communication rate, radar estimation rate, and computing energy efficiency in the considered UAV-ISCC system. 
Design of the weighted optimization between the three performance presents the challenge that the performance metrics have different units and potentially different magnitudes.
Inspired by \cite{DanQ2026TVT, XuY2023WCL}, 
${\bar{\mathcal{R}}}^{\mathrm{sec}} $, 
${\bar{\mathcal{R}}^{\mathrm{sen}}}$, 
and 
${{\bar \Phi}}$ 
are normalized with their respective bound (${\tilde{\mathcal{R}}^{\mathrm{sec}}}$, 
${\tilde{\mathcal{R}}^{\mathrm{sen}}}$, 
and 
$\tilde{{\Phi}}$) obtained in $\mathcal{P}_{1.0}$, $\mathcal{P}_{2.0}$, and $\mathcal{P}_{3.0}$ respectively.
Then, the formulated problem is expressed as
\begin{subequations}
\begin{align}
	\mathcal{P}_{4.0}: \quad &\max_{\bm{\theta}, \mathbf{F}, \tilde{\mathbf{W}}, \mathbf{Q}_s, \mathbf{Z}} \; \Xi 
	 \\
	\text{s.t.} \quad & \mathrm{(\ref{scheduling1a})-(\ref{scheduling2b}), (\ref{Opt10c})-(\ref{Opt10m}), (\ref{P1.1b})}. \nonumber
\end{align}
\end{subequations}
where 
$\Xi  = \omega_1 \frac{\bar{\mathcal{R}}^{\mathrm{sec}}}{\tilde{\mathcal{R}}^{\mathrm{sec}}} + 
\omega_2 \frac{\bar{\mathcal{R}}^{\mathrm{sen}}}{\tilde{\mathcal{R}}^{\mathrm{sen}}} + 
\omega_3 \frac{{\bar \Phi}}{\tilde{{\Phi}}}$,  
and 
$\omega_1 $, $ \omega_2$, and $ \omega_3 $ are the weighting coefficients with $\omega_1 + \omega_2 + \omega_3 = 1$.

\subsection{Communication and Sensing Scheduling Optimization}

In this subsection, we optimize the scheduling variables $\bm{\theta}$ with $\{\mathbf{F}, \tilde{\mathbf{W}}, \mathbf{Q}_s, \mathbf{Z}\}$ fixed. The optimization problem is reformulated as
\begin{subequations}
\begin{align}
	\mathcal{P}_{4.1}: \quad &\max_{\bm{\theta}} \; \bar{\omega}_1 {\bar{\mathcal{R}}}^{\mathrm{sec}} 
	+ \bar{\omega}_2 {\bar{\mathcal{R}}^{\mathrm{sen}}} 
	+ \bar{\omega}_3 {{\bar \Phi}} \\
	\text{s.t.} \quad & \mathrm{\mathrm{(\ref{scheduling1a}), (\ref{scheduling2a})}, (\ref{Opt10d}), \mathrm{(\ref{P1.1b})}-\mathrm{(\ref{scheduling2b2}), (\ref{P1.1C2})}}. \nonumber
\end{align}
\end{subequations}
where $\bar{\omega}_1 = \omega_1 / {\tilde{\mathcal{R}}^{\mathrm{sec}}}$, 
$\bar{\omega}_2 = \omega_2 / {\tilde{\mathcal{R}}^{\mathrm{sen}}}$, $\bar{\omega}_3 = \omega_3 / \tilde{{\Phi}}$. 
$\mathcal{P}_{4.1}$ is convex and can be solved using CVX.

\subsection{Computation and Power Allocation Optimization}

In this subsection, we optimize the computational frequency $\mathbf{F}$ and the beamforming matrices $\tilde{\mathbf{W}}$ with $\{\bm{\theta}, \mathbf{Q}_s, \mathbf{Z}\}$ fixed. Similar to $\mathcal{P}_{1.2}$, by dropping the rank-one constraint, $\mathcal{P}_{4.2}$ is reformulated as
\begin{subequations}
\begin{align}
	\mathcal{P}_{4.2}: \quad &\max_{\mathbf{F}, \tilde{\mathbf{W}}, \bm{\Lambda}} \; \bar{\omega}_1 {\bar{\mathcal{R}}}^{\mathrm{sec}} + \bar{\omega}_2 {\bar{\mathcal{R}}^{\mathrm{sen}}} + \bar{\omega}_3 \Psi \\
	\text{s.t.} \quad & \mathrm{(\ref{Opt10e}), (\ref{Opt10f}), (\ref{Opt10h}), (\ref{Opt10i}), (\ref{P1.1C2})}, \nonumber \\ 
	& \mathrm{(\ref{P1.2C1}), (\ref{P1.2C3}), (\ref{C16}), (\ref{C15})}. \nonumber
\end{align}
\end{subequations}
The same SCA procedure employed in solving $\mathcal{P}_{1.2}$, $\mathcal{P}_{2.2}$, and $\mathcal{P}_{3.2}$ is adopted to transform $\mathcal{P}_{4.2}$ into a convex form, which can then be efficiently solved using CVX.
Subsequently, Gaussian randomization can be applied to recover a rank-one solution.

\subsection{Horizontal Trajectory Optimization of $S$}
In this subsection, we optimize the horizontal trajectory $\mathbf{Q}_s$ with $\{\bm{\theta}, \mathbf{F}, \tilde{\mathbf{W}}, \mathbf{Z}\}$ fixed. Accordingly, $\mathcal{P}_{4.3}$ is reformulated as
\begin{subequations}
\begin{align}
	\mathcal{P}_{4.3}: \quad &\max_{\mathbf{Q}_s, \mathbf{p}^{\mathrm{h}}, \bm{\varphi}^{\mathrm{h}}, \mathbf{v}, \bm{\Lambda}} \; \bar{\omega}_1 {\bar{\mathcal{R}}}^{\mathrm{sec}} + \bar{\omega}_2 {\bar{\mathcal{R}}^{\mathrm{sen}}} + \bar{\omega}_3 \Psi \\
	\text{s.t.} \quad & \mathrm{(\ref{Opt10j}), (\ref{Opt10m}), (\ref{C1a1h}), (\ref{C1a4h}), (\ref{C1a5h})}, \nonumber \\ 
	& \mathrm{(\ref{C1a2h})-(\ref{C1a6h}), (\ref{C1a7h}), (\ref{C1a8h})}, \nonumber \\ 
	& \mathrm{(\ref{C2rq}) - (\ref{v1}), (\ref{v2t}), (\ref{C1rq}), (\ref{C2rq2}), (\ref{C16}), (\ref{C15})}. \nonumber
\end{align}
\end{subequations}
The same SCA procedure employed in solving $\mathcal{P}_{1.3}$, $\mathcal{P}_{2.3}$, and $\mathcal{P}_{3.3}$ is adopted to transform $\mathcal{P}_{4.3}$ into a convex form, which can then be efficiently solved using CVX.

\subsection{Vertical Trajectory Optimization of $S$}
In this subsection, we optimize the vertical trajectory $\mathbf{Z}$ with $\{\bm{\theta}, \mathbf{F}, \tilde{\mathbf{W}}, \mathbf{Q}_s\}$ fixed. The optimization problem is expressed as
\begin{subequations}
\begin{align}
	\mathcal{P}_{4.4}: \quad &\max_{\mathbf{Z}, \mathbf{p}^{\mathrm{v}}, \bm{\varphi}^{\mathrm{v}}, \bm{\Lambda}} \; \bar{\omega}_1 {\bar{\mathcal{R}}}^{\mathrm{sec}} + \bar{\omega}_2 {\bar{\mathcal{R}}^{\mathrm{sen}}} + \bar{\omega}_3 \Psi \\
	\text{s.t.} \quad & \mathrm{(\ref{Opt10i}), (\ref{Opt10k})-(\ref{Opt10m}), (\ref{C1a1v})-(\ref{C3a1v})}, \nonumber \\
	& \mathrm{(\ref{C3rz}), (\ref{C1rz}), (\ref{C2rq2}), (\ref{C16}), (\ref{C15})}. \nonumber
\end{align}
\end{subequations}
The same SCA procedure employed in solving $\mathcal{P}_{1.4}$, $\mathcal{P}_{2.4}$, and $\mathcal{P}_{3.4}$ is adopted to transform $\mathcal{P}_{4.4}$ into a convex form, which can then be efficiently solved using CVX.

Finally, an AO-based algorithm is developed to solve $\mathcal{P}_{4.0}$ by iteratively solving the subproblems discussed above. The overall procedure is summarized in \textbf{Algorithm 2}.

\section{Numerical Results and Analysis}
\label{sec:Simulation}

\begin{table}[tb]
	\caption{{Simulation Parameters}}
	\begin{center}
		\renewcommand{\arraystretch}{1}
		\scalebox{1}{\begin{tabular}{c| c | c| c }
				\Xhline{1.2pt}
				\textbf{Notation}   	& \textbf{Value}  & \textbf{Notation}   	& \textbf{Value}  \\
				\hline
				${V_{\max }^{\rm h}},{V_{\max }^{\rm z}} $ 	&  $20, 10$ m/s  & ${\rm B}$   & $10^6$ Hz\\
				\hline
				${z_{\min }},{z_{\max }}$ 	&  $50 , 150$ m   &  ${D_J}$   & $1$ Mbit \\
				\hline
				${T} ,{\delta _t}$		& $100, 1$ s    &  ${F_s}$  &  $1000$ cycles/bit \\
				\hline
				${C , D}$		& $11.95 , 0.14 $  &   ${f_{\min}, f_{\max}}$ &   $0.01$, $1$ GHz \\
				\hline
				${M}$ 			& $4*4$    	& ${E_{\max }^{\rm{UAV}}}$ &  $40000$ J\\
				\hline
				${\sigma _c^2,\sigma _s^2}$ 	& $- 110, -130$ dBm   & ${P_b, P_i}$  & $79.86 , 88.63$ W \\
				\hline
				${\kappa}$	& $10^{-26}$ &  ${U_{\rm tip}, {v_0}}$& $120, 4.03$ m/s\\
				\hline
				${\beta _0}$ 	& $-30$ dB  & $A$ & $0.503$ ${{\rm m}^2}$ \\
				\hline
				${\sigma _{\rm pre}^2}$ & ${-70}$ dBm &  ${d_0}$  & $0.6$ \\
				\hline
				 ${ \delta, {\hat \gamma}  } $ 	& $ 0.01, \pi /\sqrt 3 $&  $\rho $ & $1.225$ kg/${{\rm m}^3}$ \\
				\hline
				$\mu$ 	& $2 \times 10^{-5}$ s  &  $s$  & $0.05$ ${{\rm m}^3}$ \\
				\hline
				$\alpha_1, \alpha_2 , \alpha_3$ & 10, 5, 1 & $\xi$ & 1 ${\rm m}^2$ \\
				\hline
				${\epsilon}$  &   $0.01$   & $W$ & $20$ $N$ \\
				\Xhline{1.2pt}
		\end{tabular}}
	\end{center}
	\label{table2}
\end{table}

In this section, numerical results are presented to evaluate the performance of the proposed algorithm. The four ground communication users are located at $[250, 50]^T$, $[450, 150]^T$, $[300, 400]^T$, and $[50, 300]^T$, respectively, while the four sensing targets are positioned at $[250, 100]^T$, $[350, 150]^T$, $[325, 300]^T$, and $[150, 270]^T$. $B$ and $E$ are located at $[250, 250]^T$ m and $[300, 200]^T$, respectively. All nodes are distributed within a $500 \times 500$ m$^2$ ground area, and $S$’s starting and ending points are set at $[100, 100]^T$ with an initial flight altitude of $50$ m. The remaining parameter configurations are summarized in Table \ref{table2}.
The communication-centric, sensing-centric, computation-centric, and weighted optimization for communication, sensing, and computing are denoted as ``Com", ``Sen", ``Cop", and ``CSC", respectively. For the CSC design, the weighting coefficients are set to $\omega_1 = \omega_2 = \omega_3 = 1/3$.

Figs. \ref{fig02a}-\ref{fig02f} present the optimized parameters and performance with different schemes. Specifically, Figs. \ref{fig02a} and \ref{fig02b} show the 3D and 2D trajectories, respectively, while Fig. \ref{fig02c} illustrates the flight altitude. From these figures, it can be observed that in Com, which aims to achieve optimal communication performance, $S$ visits the communication users in a sequential manner; when flying toward the next user, it increases its altitude to obtain a higher LoS probability, reflecting \textit{the trade-off between LoS probability and distance} as noted in Ref. \cite{LeiH2024TCCN3D}. In Sen, which targets optimal sensing performance, $S$ sequentially visits the ground targets and maintains the minimum flight altitude for most time slots. This represents \textit{a key difference between Com and Sen}, as shown in Eqs. (\ref{hc}) and (\ref{hvj}): \textit{in the sensing channel, the SINR is inversely proportional to the fourth power of distance, making distance a more dominant factor than LoS probability in determining sensing performance, thus requiring closer proximity to maximize sensing performance.} In Cop, computational performance is prioritized, prompting $S$ to fly as close as possible to $B$ to offload data and reduce energy consumption; for most time slots, the variations in both horizontal and vertical positions are significantly smaller than those in the Com and Sen schemes. In CSC, where communication, sensing, and energy efficiency are jointly considered, the horizontal and vertical trajectories lie between those of the Com and Sen schemes. Figs. \ref{fig02d} and \ref{fig02e} depict the optimized horizontal and vertical velocities, respectively. As shown in Fig. \ref{fig02d}, in Com, $S$ flies at maximum speed between communication users and hovers above users to maximize the communication rate; similarly, in Sen, it flies at maximum speed between sensing targets and hovers above the scheduled targets to maximize the sensing rate. In CSC, where both communication and sensing are considered, $S$ hovers in regions close to both communication users and sensing targets. It can also be observed from Fig. \ref{fig02d} that with Com and CSC, $S$ hovers for a longer duration $U_4$ at a relatively higher altitude, as $U_4$ is farthest from $E$, thereby enabling better communication performance. As shown in Figs. \ref{fig02d} and \ref{fig02e}, in Cop, both horizontal and vertical velocities are kept low to conserve energy. Finally, Fig. \ref{fig02f} illustrates the data offloading and local computing capabilities of $S$ in Cop; since $S$ remains in close proximity to $B$ for most time slots, a strong offloading capability is achieved.

To analyze the trade-off among the three types of performance, we compare the results of the CSC scheme in Fig. \ref{fig03} corresponding to the following four sets of weighting coefficients, defined as ``Uniform" ($\omega_1 = \omega_2 = \omega_3 = 1/3$), ``Com\_Max" ($\omega_1 = 0.8, \omega_2 = \omega_3 = 0.1$), ``Sen\_Max" ($\omega_1 = 0.1, \omega_2 = 0.8, \omega_3 = 0.1$), and ``Cop\_Max" ($\omega_1 = \omega_2 = 0.1, \omega_3 = 0.8$), respectively. Specifically, Figs. \ref{fig03a} and \ref{fig03b} show the 3D and 2D trajectories, respectively, while Fig. \ref{fig03c} illustrates the flight altitude. It can be observed that in Fig. \ref{fig03a}, Cop\_Max exhibits a significant difference from Cop in Fig. \ref{fig02b}, as Cop\_Max jointly maximizes both communication and sensing performance, resulting in a horizontal trajectory similar to those of Com and CSC. Moreover, the distinction between Fig. \ref{fig02c} and Fig. \ref{fig03c} is pronounced, as the latter requires simultaneous optimization of all three performance metrics. Specifically, compared with Com, Com\_Max additionally accounts for energy efficiency, leading to a significantly lower flight altitude, while compared with Cop, Cop\_Max incorporates communication performance, resulting in a higher flight altitude. Figs. \ref{fig03d} and \ref{fig03e} depict the horizontal and vertical velocities under different weight configurations, respectively, showing that Cop\_Max exhibits higher velocities than Cop to maximize overall performance, whereas Com\_Max shows lower velocities than Com to conserve energy. Finally, Fig. \ref{fig03f} illustrates the data offloading and local computing capabilities under Cop\_Max, where communication and sensing performances are maximized while maintaining offloading capability in the former half. In the latter half of the trajectory, $S$ flies at a higher altitude, increasing the LoS probability between $S$ and $B$ and thereby enhancing offloading capability.

Figs.  \ref{fig04a} and \ref{fig04b} illustrate the scheduling results for communication users and sensing targets for Uniform, respectively, where it can be observed that users (targets) are scheduled in turn. Notably, no user is scheduled in the 51-st time slot, as $S$ flies too close to $E$, failing to meet the requirements for secure communication. Figs.  \ref{fig04c} - \ref{fig04f} present the 2D and 3D views of the normalized beamforming gain with Uiniform for the 18-th and 52-nd time slots. Specifically, in the 18-th time slot, the beamforming gain is concentrated on $U_1$, with some gain also observed at $V_1$ and $B$. In the 52-nd time slot, the beamforming gain is primarily focused on $U_3$ and $V_3$, with a small portion distributed to $E$. It can be seen that in the 18-th time slot, the scheduled user is $U_1$ and the scheduled target is $V_1$, while in the 52-nd time slot, the scheduled is $U_3$ and the scheduled target is $V_3$.

\section{Conclusion}
\label{sec:Conclusion}
  
This work investigated a secure UAV-ISCC system, wherein the UAV provides communication services to ground users, performs sensing for ground targets, and either processes the sensed data locally or offloads it to a ground base station. First, by jointly optimizing the UAV's 3D trajectory, beamforming, communication and sensing scheduling, and computational frequency, the secure communication rate, radar estimation rate, and computational energy efficiency were maximized, respectively. Subsequently, the normalized weighted trade-off among the three performances was optimized. Numerical results demonstrate the convergence and effectiveness of the proposed algorithm. 



\begin{thebibliography}{1}

\bibitem{ZhangD2026Surveys}	
D. Zhang, Y. Cui, X. Cao, N. Su, Y. Gong, F. Liu, W. Yuan, X. Jing, J. Andrew Zhang, J. Xu, C. Masouros, D. Niyato, and M. Di Renzo, ``Integrated sensing and communications over the years: An evolution perspective,'' \textit{IEEE Communications Surveys \& Tutorials}, vol. 28, pp. 5014-5048, Mar. 2026.

\bibitem{WangX20024ACS} 
X. Wang, Q. Guo, Z. Ning, L. Guo, G. Wang, X. Gao, and Y. Zhang, ``Integration of sensing, communication, and computing for metaverse: A survey,'' \textit{Acm Comput. Surv.}, vol. 56, no. 10, pp. 1-38, May 2024.

\bibitem{LiC2026Surveys}	
C. Li, M. Dong, Y. Fu, F. Richard Yu, and N. Cheng, ``Integrated sensing, communication, and computation for IoV: Challenges and opportunities,'' \textit{IEEE Commun. Surveys Tuts.}, vol. 28, pp. 1136-1168, Jan. 2026.
	
\bibitem{WenD2025CST} 
D. Wen, Y. Zhou, X. Li, Y. Shi, K. Huang, and K. B. Letaief, ``A survey on integrated sensing, communication, and computation,'' \textit{IEEE Commun. Surveys Tuts.}, vol. 27, no. 5, pp. 3058-3098, Oct. 2025.

\bibitem{ZhaoY2025TCOM} 
Y. Zhao, Q. Wu, W. Chen, Y. Zeng, R. Liu, W. Mei, F. Hou, and S. Ma, ``Multi-functional beamforming design for integrated sensing, communication, and computation,'' \textit{IEEE Trans. Commun.}, vol. 73, no. 8, pp. 6322-6336, Dec. 2025.

\bibitem{WangZ2023JSAC} 
Z. Wang, X. Mu, Y. Liu, X. Xu, and P. Zhang, ``NOMA-aided joint communication, sensing, and multi-tier computing systems,'' \textit{IEEE J. Sel. Areas Commun.}, vol. 41, no. 3, pp. 574-588, Mar. 2023.

\bibitem{CangY2025JSAC} 
Y. Cang, M. Chen, and Z. Yang, ``Cooperative detection for MEC-aided multi-static ISAC systems,'' \textit{IEEE J. Sel. Areas Commun.}, vol. 44, pp. 545-561, Feb. 2026.

\bibitem{DingY2023JSTSP} 
Y. Ding, Y. Feng, W. Lu, S. Zheng, N. Zhao, L. Meng, A. Nallanathan, and X. Yang, ``Online edge learning offloading and resource management for UAV-assisted MEC secure communications,'' \textit{IEEE J. Sel. Topics Signal Process.}, vol. 17, no. 1, pp. 54-65, Jan. 2023.


\bibitem{DingC2022JSAC} 
C. Ding, J.-B. Wang, H. Zhang, M. Lin, and G. Y. Li, ``Joint MIMO precoding and computation resource allocation for dual-function radar and communication systems with mobile edge computing,'' \textit{IEEE J. Sel. Areas Commun.}, vol. 40, no. 7, pp. 2085-2102, Jul. 2022.


\bibitem{SunG2024TVT} 
G. Sun, X. Wu, W. Hao, Z. Zhu, J. Li, Z. Chu, and P. Xiao, ``Resource management for integrated communications, computing, and sensing (ICCS) networks,'' \textit{IEEE Trans. Veh. Technol.}, vol. 73, no. 12, pp. 18719-18731, Dec. 2024.

\bibitem{WangY2025TCCN} 
Y. Wang, G. Sun, Z. Sun, J. Wang, J. Li, C. Zhao, J. Wu, S. Liang, M. Yin, P. Wang, D. Niyato, S. Sun, and D. In Kim, ``Toward realization of low-altitude economy networks: Core architecture, integrated technologies, and future directions,'' \textit{IEEE Trans. Cogn. Commun. Netw.}, vol. 11, no. 5, pp. 2788-2820, Apr. 2025.

\bibitem{ChenJ2024WCL} 
J. Chen, Y. Xu, D. Yang, and T. Zhang, ``UAV-assisted ISCC networks: Joint resource and trajectory optimization,'' \textit{IEEE Wireless Commun. Lett.}, vol. 13, no. 9, pp. 2372-2376, Sep. 2024.


\bibitem{ZhouY2024IoT}
Y. Zhou, X. Liu, X. Zhai, Q. Zhu, and T. S. Durrani, ``UAV-enabled integrated sensing, computing, and communication for Internet of things: Joint resource allocation and trajectory design,'' \textit{IEEE Internet Things J.}, vol. 11, no. 7, pp. 12717-12727, Apr. 2024.

\bibitem{XuY2023WCL} 
Y. Xu, T. Zhang, Y. Liu, and D. Yang, ``UAV-enabled integrated sensing, computing, and communication: A fundamental trade-off,'' \textit{IEEE Wireless Commun. Lett.}, vol. 12, no. 5, pp. 843-847, May 2023.

\bibitem{VanChienT2024CL}
T. V. Chien, M. D. Cong, N. C. Luong, T. N. Do, D. I. Kim, and S. Chatzinotas, ``Joint computation offloading and target tracking in integrated sensing and communication enabled UAV networks,'' \textit{IEEE Commun. Lett.}, vol. 28, no. 6, pp. 1327-1331, Jun. 2024.

\bibitem{XuS2025TMC}
S. Xu, Z. Liu, L. Zhao, Z. Liu, X. Wang, Z. Fei, and A. Nallanathan, ``Joint trajectory and beamforming optimization for AAV-relayed integrated sensing and communication with mobile edge computing,'' \textit{IEEE Trans. Mobile Comput.}, vol. 24, no. 10, pp. 11180-11192, May 2025.


\bibitem{ZhouY2026TGCN}
Y. Zhou and X. Liu, ``Trade-off between radar sensing and energy consumption in integrated sensing, computing, and communication UAV network,'' \textit{IEEE Trans. Green Commun. Netw.}, vol. 10, pp. 511-521, Jan. 2026.

\bibitem{HuangN2023IoT} 
N. Huang, C. Dou, Y. Wu, L. Qian, B. Lin, and H. Zhou, ``Unmanned-aerial-vehicle-aided integrated sensing and computation with mobile-edge computing," \textit{IEEE Internet Things J.}, vol. 10, no. 19, pp. 16830-16844, Oct. 2023.

\bibitem{PengS2025TCOM} 
S. Peng, B. Li, L. Liu, Z. Fei, and D. Niyato, ``Trajectory design and resource allocation for multi-UAV-assisted sensing, communication, and edge computing integration,'' \textit{IEEE Trans. Commun.}, vol. 73, no. 4, pp. 2847-2861, Apr. 2025.



\bibitem{LeiH2025IoTCongke}
H. Lei, C. Jiang, K.-H. Park, M. A. Aboulhassan, S. Zhou, and G. Pan, ``On secure UAV-aided ISCC systems,'' \textit{IEEE Internet Things J.}, vol. 12, no. 19, pp. 40851-40862, Oct. 2025.

\bibitem{ZhangQ2026JSAC}
Q. Zhang, \textit{et al.}, ``Covert transmission for active RIS-aided full-duplex UAV integrated sensing, communication, and computation systems," \textit{IEEE J. Sel. Areas Commun.}, vol. 44, pp. 1992-2007, Feb. 2026.

\bibitem{AlHouraniA20217WCL}
A. Al-Hourani, S. Kandeepan, and S. Lardner, ``Optimal LAP altitude for maximum coverage," \textit{IEEE Wireless Commun. Lett.}, vol. 3, no. 6, pp. 569-572, Dec. 2014.

\bibitem{LeiH2025TCCN} 
H. Lei, D. Meng, H. Ran, K.-H. Park, G. Pan, and M.-S. Alouini, ``Multi-UAV trajectory design for fair and secure communication,'' \textit{IEEE Trans. Cogn. Commun. Netw.}, vol. 11, no. 3, pp. 1966-1980, Jun. 2025.


\bibitem{DengC2023TWC} 
C. Deng, X. Fang, and X. Wang, ``Beamforming design and trajectory optimization for UAV-empowered adaptable integrated sensing and communication," \textit{IEEE Trans. Wireless Commun.}, vol. 22, no. 11, pp. 8512-8526, Nov. 2023.

\bibitem{LiuF2018TSP} 
 F. Liu, C. Masouros, A. Li, T. Ratnarajah, and J. Zhou, ``MIMO radar and cellular coexistence: A power-efficient approach enabled by interference exploitation,'' \textit{IEEE Trans. Signal Process.}, vol. 66, no. 14, pp. 3681-3695, Jul. 2018.
 
\bibitem{JingX2024TWC} 
X. Jing, F. Liu, C. Masouros, and Y. Zeng, ``ISAC from the sky: UAV trajectory design for joint communication and target localization,'' \textit{IEEE Trans. Wireless Commun.}, vol. 23, no. 10, pp. 12857-12872, Oct. 2024.

\bibitem{MengK2023TWC}
K. Meng, Q. Wu, S. Ma, W. Chen, K. Wang, and J. Li, ``Throughput maximization for UAV-enabled integrated periodic sensing and communication,'' \textit{IEEE Trans. Wireless Commun.}, vol. 22, no. 1, pp. 671-687, Jan. 2023.

\bibitem{ZhangJ2024TWC}  
J. Zhang, J. Xu, W. Lu, N. Zhao, X. Wang, and D. Niyato, ``Secure transmission for IRS-aided UAV-ISAC networks," \textit{IEEE Trans. Wireless Commun.}, vol. 23, no. 9, pp. 12256-12269, Sep. 2024.

\bibitem{LeiH2024TCCN3D} 
H. Lei, X. Wu, K.-H. Park, and G. Pan, ``3D trajectory design for energy-constrained aerial CRNs under probabilistic LoS channel,'' \textit{IEEE Trans. Cogn. Commun. Netw.}, vol. 11, no. 3, pp. 1522-1534, Jun 2025.

\bibitem{ChiriyathARTSP2016}
A. R. Chiriyath, B. Paul, G. M. Jacyna, and D. W. Bliss, ``Inner bounds on performance of radar and communications co-existence," \textit{IEEE Trans. Signal Process.}, vol. 64, no. 2, pp. 464-474, Sep. 2016.

\bibitem{LuW2022TCOM}
W. Lu, et al., ``Secure NOMA-based UAV-MEC network towards a flying eavesdropper," \textit{IEEE Trans. Commun.}, vol. 70, no. 5, pp. 3364-3376, May 2022.


\bibitem{ZengY2019TWC} 
Y. Zeng, J. Xu, and R. Zhang, ``Energy minimization for wireless communication with rotary-wing UAV,'' \textit{IEEE Trans. Wireless Commun.}, vol. 18, no. 4, pp. 2329-2345, Apr. 2019.

\bibitem{LeiH2025TAES} 
H. Lei, D. Meng, K.-H. Park, N. Saeed, and G. Pan, ``DRL-based resource allocation for aerial IoT systems with no-fly-zones,'' \textit{IEEE Trans. Aerosp. Electron. Syst.}, vol. 61, no. 6, pp. 17892 - 17905, Dec. 2025.

\bibitem{DanQ2026TVT}
Q. Dan, H. Lei, K.-H. Park, G. Pan, and M.-S. Alouini, ``Two birds with one stone: Beamforming design for joint target sensing and proactive eavesdropping," \textit{IEEE Trans. Veh. Technol.}, early access, doi: 10.1109/TVT.2026.3679570, Mar. 2026.

\bibitem{WangW2022JSAC} 
W. Wang, W. Ni, H. Tian, Z. Yang, C. Huang, and K.-K. Wong, ``Safeguarding NOMA networks via reconfigurable dual-functional surface under imperfect CSI,'' \textit{IEEE J. Sel. Topics Signal Process.}, vol. 16, no. 5, pp. 950-966, Aug. 2022.




\end{thebibliography}
\end{document}